\documentclass[10pt,journal,twocolumns]{IEEEtran}
\usepackage{subfigure}
\usepackage{amssymb}
\usepackage{amsthm}
\usepackage{amsmath}
\usepackage{mathrsfs}
\usepackage{hyperref}
\usepackage{graphicx}
\usepackage{colortbl,dcolumn}
\usepackage{booktabs}
\usepackage{xcolor}
\usepackage{cite}
\usepackage{threeparttable}
\usepackage[most]{tcolorbox}
\usepackage{arydshln}
\usepackage{algorithm}
\usepackage{algorithmic}
\usepackage{blkarray}
\usepackage{cases}

\usepackage{stmaryrd}

\newtheorem{definition}{Definition}
\newtheorem{lemma}{Lemma}
\newtheorem{theorem}{Theorem}
\newtheorem{proposition}{Proposition}
\newtheorem{corollary}{Corollary}
\newtheorem{remark}{Remark}

\newcommand{\defref}[1]{Definition~\ref{#1}}
\newcommand{\lemref}[1]{Lemma~\ref{#1}}
\newcommand{\thmref}[1]{Theorem~\ref{#1}}
\newcommand{\propref}[1]{Proposition~\ref{#1}}
\newcommand{\corref}[1]{Corollary~\ref{#1}}
\newcommand{\figref}[1]{Fig.~\ref{#1}}

\newcommand{\secref}[1]{Section~\ref{#1}}
\newcommand{\apxref}[1]{Appendix~\ref{#1}}

\newcommand{\algref}[1]{Algorithm~\ref{#1}}

\usepackage{mathtools} % \DeclareDelimiter etc
\newenvironment{Proof}{\noindent{\em Proof:\/}}{\hfill $\blacksquare$\par}

\usepackage{enumerate}
\ifCLASSOPTIONcompsoc
\else
\fi
\ifCLASSINFOpdf
\else
\fi
\graphicspath{{pdfs/},{images/}}

\begin{document}
		
\title{Asymptotic Boundedness of Distributed Set-Membership Filtering}
				\author{Yudong~Li,~Yirui~Cong,~Shimin~Wang, Martin~Guay,~Jiuxiang~Dong
                \IEEEcompsocitemizethanks{\IEEEcompsocthanksitem Y.~Li and J.~Dong are with the State Key Laboratory of Synthetical Automation of Process Industries, Northeastern University, China, (email: 2210325@stu.neu.edu.cn; dongjiuxiang@ise.neu.edu.cn).\protect\\
                \indent Y.~Cong is with the College of Intelligence Science and Technology, National University of Defence Technology, China (email: congyirui11@nudt.edu.cn).
                % \indent  is with the Massachusetts Institute of Technology, Cambridge, USA (e-mail: bellewsm@mit.edu).
                \indent S.~Wang and M.~Guay are with Queen's University, Kingston, ON K7L 3N6, Canada (e-mail: guaym@queensu.ca).
                }
                }% <-this % stops a space
		
\IEEEtitleabstractindextext{%
\begin{abstract}
            Asymptotic boundedness is a crucial property of Distributed Set-Membership Filtering (DSMFing) that prevents the unbounded growth of the set estimates caused by the wrapping effect. 
            However, this important property remains underinvestigated, compared to its noise-free and stochastic-noise counterparts, i.e., the convergence of Distributed Observers (DOs) and the bounded error covariance of Distributed Kalman Filters (DKFs). 
            This paper studies the asymptotic boundedness of DSMFing for linear discrete-time systems.
            A novel concept, termed the Collective Observation-Information Tower (COIT), is introduced to characterize the fundamental relationship between the structure of graphs and the set estimates, which enables the boundedness analysis. 
            Leveraging the COIT, an easily verifiable sufficient condition for the asymptotic boundedness of linear DSMFing is established. 
            Surprisingly, the sufficient condition generalizes the well-known collective detectability condition for DOs and DKFs; it links DSMFs to existing distributed estimation methods and reveals the unique characteristic of DSMFs.
            %The effectiveness of the theoretical results is validated through simulations.
		\end{abstract}
		
		}

	\maketitle
		\IEEEdisplaynontitleabstractindextext
		\IEEEpeerreviewmaketitle
	\section{Introduction}\label{sec:Introduction}

Distributed Set-Membership Filters (DSMFs) serve as an effective solution to handle distributed state estimation problems when systems are corrupted by unknown-but-bounded noises~\cite{8425647,cong2022stability,10750409}.  
Thus, the DSMFs have great potential for broad applications, e.g., in sensor networks of autonomous vehicles~\cite{garcia2020guaranteed}. %and cyber-physical systems~\cite{LI202149}.

In the literature, there are two main classes of DSMFs w.r.t. the set representations, shown as follows:
    %	
    % Similar to all kinds of distributed state estimation methods, the bounds of the set estimate from DSMFing need to be guaranteed, which shows its effectiveness.
    %  %
    % To improve the filtering performance of DSMFs, the present research utilizes different optimization strategies on the set estimate of DSMFs, which is illustrated in terms of different set representations as follows:
    \begin{itemize}
        \item \textbf{Zonotopic DSMF:}
            Zonotopic DSMFs aim to determine the optimal observer gain that minimizes the $F$-radius of the fused zonotopic set estimate.
            For instance, \cite{orihuela2017distributed} established a zonotopic DSMF for interconnected multi-rate linear systems, where the optimal observer gain minimizing the $F$-radius of the filtering zonotope was obtained analytically. 
            Building on this work, \cite{ierardi2021distributed} introduced a multi-hop staircase decomposition into the zonotopic DSMF framework to reduce computational complexity under the collectively detectable condition, while also deriving an analytical solution for the optimal observer gain using a similar strategy.
            Furthermore, employing a comparable methodology, \cite{ZHU2025112345} designed a mean-shift-based robust zonotopic DSMF to handle outlier contamination in DSMFing. 
            In addition, \cite{alanwar2023distributed} proposed two DSMFs based on strip and set-propagation approaches, where the optimal parameters of the over-approximating zonotopic set estimate were obtained analytically.
            
            \item \textbf{Ellipsoidal DSMF:}
            Similar to the zonotopic DSMFs, ellipsoidal DSMFs for linear systems focus on deriving optimal observer gains to minimize the fused ellipsoidal estimate. 
            For example, \cite{Liu2021} employed a coding–decoding communication strategy to reduce the communication burden in time-varying sensor networks, where the observer gain was obtained by solving an LMI-based optimization problem minimizing the trace of the ellipsoidal estimate. 
            In \cite{XIE2025112347}, an ellipsoidal DSMF was designed for privacy-preserving purposes in linear systems, where an analytical observer gain minimized the trace of the ellipsoidal estimate. 
            For nonlinear systems, \cite{8793114} introduced an event-triggered transmission scheme to handle resource constraints, with the optimal parameter recursively computed through an LMI-based optimization problem to yield ellipsoids of minimum trace.
            Moreover, \cite{8605375} developed a DSMF for multirate systems under Round-Robin scheduling; instead of minimizing the trace, the observer gain was obtained from an LMI-based optimization that enforces an optimal finite-horizon constraint.
    \end{itemize}
    Overall, although zonotopic and ellipsoidal DSMFs provide analytical or optimization-based solutions for minimizing set estimates, they still suffer from the wrapping effect, which can lead to unbounded growth of the set estimates.
    Indeed, it is unclear when the estimates can be guaranteed to stay bounded as time elapses.
    %, even for zonotopic and ellipsoidal DSMFs.
    %
    % In fact, regardless of the chosen set representation, existing DSMFs remain susceptible to the wrapping effect, and the boundedness property—despite its fundamental importance for ensuring stability and reliability in long-term applications—has thus far received limited attention in the literature. 
    %  
    
    It should be pointed out that other distributed state estimation methods, such as Distributed Observers (DOs) for noise-free systems and Distributed Kalman Filters (DKFs) for systems with Gaussian noises, have achieved comprehensive and unified conclusions for convergence/boundedness analysis.
    For linear discrete-time systems, in \cite{7463019}, a necessary and sufficient condition for the asymptotic convergence of DOs was proposed based on the consensus of the Luenberger observer, which is collectively detectable.
    The same condition (collective detectability) was also established for continuous-time systems in \cite{wang2024distributed}.
    This condition also provided a criterion for analyzing the asymptotic boundedness of the error covariance in DKFs for stochastic systems.
    In \cite{BATTISTELLI2014707}, supported by the fused Kullback–Leibler average of the local probability density functions, the bounds of error covariance of DKFs were guaranteed under the collective detectability condition for linear time-invariant systems.
    A similar boundedness condition for DKFs, known as collectively uniformly detectable, was also achieved in linear time-varying systems \cite{8845692}.

    Unlike those two distributed state estimation methods (i.e., DOs and DKFs), no effective approaches have been developed to analyze the boundedness of DSMFing.
    %leaving the issue of boundedness analysis of DSMFing inadequately addressed.
    %
    To the best of our knowledge, explicit conditions that guarantee bounded set estimates of DSMFing are still lacking, even for zonotopic and ellipsoidal DSMFs.
    
 %    \begin{table}[H]\label{T1}
	% 	\caption{\textbf{Summary of Distributed State Estimation Method}}
	% 	\centering
	% 	\begin{tabular}{cccc}
	% 		\toprule
	% 		Reference& Method & Existence condition\\
	% 		\midrule 			\cite{7463019,mitra2018distributed,8093658}&DO & Collectively detectable\\ \cite{BATTISTELLI2014707}, \cite{8845692}&DKF& Collectively detectable\\
	% 		% &DIO&Interval&Data7\\
 %            - &DSMF&?\\			
	% 		\bottomrule
	% 	\end{tabular}
	% \end{table}
	% %
    
    Motivated by the discussions above, we aim to provide a unified analysis of boundedness for DSMFing. 
    %
    % We seek to find a more accurate way to express the data fusion in DSMFing, and develop an explicit condition that can guarantee an asymptotically bounded state estimation for DSMFing, analogous to those demonstrated for DO and DKF.
    % % .
    %
    The main contributions of this article are highlighted as follows:
    \begin{itemize}
        \item By modeling a unified DSMFing process with uncertain variables, the fusion of the local set estimates is outer bounded by a new concept called Collective Observation-Information Tower (COIT).
        The proposed COIT indicates the fundamental relationship between the structure of graphs (more specifically, the source components) and the set estimates.
        Consequently, it enables the boundedness analysis of DSMFing.
        
        \item Based on the proposed COIT, an easily checked sufficient condition for the asymptotic boundedness of linear DSMFing is established. 
        Surprisingly, the sufficient condition generalizes the well-known collective detectability condition for DOs and DKFs.
        This presents two key benefits: (i) the collective detectability is applicable to DSMFs, thereby bridging the gap between DSMFs and the existing estimation methods;
        (ii) the sufficient condition reveals the unique characteristic of DSMFing.
        % %
        % Compared to the collective detectability condition, the proposed boundedness condition is more general.
    \end{itemize}
	In the rest of the paper, the preliminaries of this study are given in \secref{sec:Prelinimaries}. The system model and the problem description of DSMFing are presented in \secref{sec:problemdescript}.
    Then, an asymptotic boundedness condition of the DSMFing problem is established in \secref{sec:sufficiency condition}.
    Finally, the proposed asymptotic boundedness condition is corroborated via a simulation example in \secref{sec_num}.
\section{Preliminaries}\label{sec:Prelinimaries}
   \subsection{Notation}
	Throughout this paper, we use $\left\|  \cdot  \right\|$ to represent the Euclidean norm (of a vector) or the spectral norm (of a matrix).
	For a sample space $\Omega$, a measurable function ${\mathbf{x}}:\Omega  \to \mathcal{X}$ from sample space $\Omega$ to measurable set $\mathcal{X}$ is called an uncertain variable~\cite{6415998,cong2021rethinking}, with its range $\llbracket\mathbf{x}\rrbracket$ defined by:
	\begin{equation}\label{eq_uncertain_varibale}
		\llbracket\mathbf{x}\rrbracket := \left\{\mathbf{x}(\omega)\colon \omega \in \Omega\right\}.
	\end{equation}
	$D(\mathcal{S})=\sup_{s,s' \in {\mathcal{S}_k}}\left\| {s - s'} \right\|$ stands for the diameter of $\mathcal{S}$. 
	For multiple uncertain variables with consecutive indices, we
	define ${{\mathbf{x}}_{{k_1}:{k_2}}} = {{\mathbf{x}}_{{k_1}}}, \ldots ,{{\mathbf{x}}_{{k_2}}}$. 
	$I_n$ and $0_{n,m}$ stand for identity and null matrices with compatible dimensions, respectively. 
	Given two sets ${\mathcal{S}_1}$ and ${\mathcal{S}_2}$ in a Euclidean space, the operation $ + $ stands for the Minkowski sum of ${\mathcal{S}_1}$ and ${\mathcal{S}_2}$. 
	The summation $\sum\nolimits_{i = a}^b {{\mathcal{S}_i}}$ represents ${\mathcal{S}_a} + {\mathcal{S}_{a + 1}} +  \cdots  + {\mathcal{S}_b}$ for $a \leqslant b$, but returns $\{0\}$ for $a>b$.\footnote{This is different from the mathematical convention but can keep many
		expressions compact in this paper.}
	Given a matrix $M \in {\mathbb{R}^{n \times m}}$, the kernel, the range space, the Moore-Penrose inverse, and the transpose are $\ker (M)$, $R(M)$, $M^\dag$, and $M^\top$, respectively;
	when $M$ is a square matrix, its spectrum is denoted by $\Lambda_M$.
        The interval hull $\overline {{\text{IH}}} (\mathcal{S})$ of a bounded set $\mathcal{S}$ is the smallest outerbounding interval of $\mathcal{S}$\cite{moore2009introduction}.
        %
        %For row vector $a$ and $b$, ${\rm col}(a,b)$ returns the matrix $[a^\top b^\top]^\top$. 
        For a set of matrices $\left\{X_i|\ i=1, 2, \dots, N\right\}$ and a set of their index $\mathcal{N}=\left\{1, 2, \cdots, N\right\}$, define ${\rm{col}}{(X_1, \dots, X_N)}$ as a matrix formed by stacking them (i.e., 
$\setlength\arraycolsep{1.4pt}
{\left[ {\begin{array}{*{20}{c}}
			{X_1^{\top}}&{X_2^{\top}}& \dots &{X_N^{\top}}
	\end{array}} \right]^{\top}}$) if dimensions matched.
        ${\rm diag}(A, B)$ stands for the diagonal matrix form by square matrix $A$ and $B$.

	% %
	% The kernel is denoted by $\ker (A)$.
	% %
	% $R(A)$ stand for the row space of matrix $A$.
	% %
	% $\Lambda_A$ stands for the set consists of all the eigenvalue of $A$.
	% %
	% Given a matrix $M \in {\mathbb{R}^{n \times m}}$, the Moore-Penrose inverse is ${M^+ }$.
	% %
	% $A^\top$ stands for the transpose of $A$.
	% %
	
	\section{System Model and Problem Description}\label{sec:problemdescript}
	Consider a discrete-time linear sampled-data system\footnote{For a sampled-data system, the system matrix is non-singular, which is widely considered in discrete-time filtering problems \cite{BATTISTELLI2014707}, \cite{8845692}.}
	\begin{equation}\label{eq_intro_system}
		{{\mathbf{x}}_{k + 1}} = A{{\mathbf{x}}_k} + B{{\mathbf{w}}_k},
	\end{equation}
	where $A \in {\mathbb{R}^{n \times n}}$ and $B \in {\mathbb{R}^{n \times p}}$. 
	The system state is $\mathbf{x}_k$ (with its realization $x_k \in \llbracket\mathbf{x}_k\rrbracket \subseteq \mathbb{R}^n$), and ${{\mathbf{w}}_k}$ stands for the process noise (with its realization $w_k \in \llbracket\mathbf{w}_k\rrbracket \subseteq \mathbb{R}^p$).
	The system state is estimated by a network of $N$ sensors.
	Each sensor $i$ in the network can obtain local measurements at time ${k \in {\mathbb{N}_0}}$, with the following measurement equation:
	\begin{equation}\label{eq_xx}
		{\mathbf{y}}_k^i = {C_i}{{\mathbf{x}}_k} + {\mathbf{v}}_k^i,
	\end{equation}
	where $C_i$ is null if sensor $i$ has no measurements; 
	${\mathbf{y}}_k^i$ and ${\mathbf{v}}_k^i$ are the measurement (with its realization $y_k^i \in \llbracket\mathbf{y}_k^i\rrbracket \subseteq \mathbb{R}^{q_i}$) and measurement noise (with its realization $v_k^i \in \llbracket\mathbf{v}_k^i\rrbracket \subseteq \mathbb{R}^{q_i}$) of sensor $i$, respectively.
	Besides, the process and measurement noises satisfy
        \begin{align}\label{eqn:Uniformly Bounded Noises}
            \mathop {\sup }\limits_{k \in {\mathbb{N}_0}} D(\llbracket {{{\mathbf{w}}_k}} 
	\rrbracket) < \infty&~~~~\textnormal{and}& \mathop {\sup }\limits_{k \in {\mathbb{N}_0}} D(\llbracket {{{\mathbf{v}}_k^i}} 
	\rrbracket) < \infty.
        \end{align}
	% \[\mathop {\sup }\limits_{k \in {\mathbb{N}_0}} D(\llbracket {{{\mathbf{w}}_k}} 
	% \rrbracket) < \infty ,\quad \mathop {\sup }\limits_{k \in {\mathbb{N}_0}} D(\llbracket {{{\mathbf{v}}_k^i}} 
	% \rrbracket) < \infty. \]
	%
	Sensors can exchange information with their neighbors in the network. 
	The communication topology is expressed by a directed graph $G = (V,E)$, where $V = \{1,\ldots,N\}$ is the set of sensors and $E \subseteq V \times V$ denotes the set of edges.
	
	For sensor $i \in V$, a DSMF aims to provide a set estimate/approximation containing all possible $x_k$ for $k\in\mathbb{N}_0$ based on its local measurements and the information obtained from its neighbors, which is called a belief $\mathcal{B}_i({{\mathbf{x}}_k})$ defined as follows.
	\begin{definition}[Belief]
		From the perspective of sensor $i$, a belief $\mathcal{B}_i({{\mathbf{x}}_k} 
		)$ derived by a DSMF at $k \in {\mathbb{N}_0}$ is an outer bound of posterior range $\llbracket\mathbf{x}_k|y_{0:k}\rrbracket$, where $\llbracket {{\mathbf{x}}_k|y_{0:k}} 
    	\rrbracket$ describes all possible states $x_k$ given all the measurements in $V$ up to $k$, i.e., ${y_{0:k}}: = {y_{0}}, \ldots ,{y_{k}}$, with ${y_k} = \textnormal{col}({{y_k^1 }},\dots {{y_k^N}})$ being the joint measurement vector of the whole network at time $k$.           
	\end{definition}
	\begin{remark}
		The estimate provided by the DSMF methods in \cite{8552364,Liu2021,9591293,8605375,orihuela2017distributed,ierardi2021distributed,8793114} can be regarded as different beliefs.
		Those beliefs are outer bounds of $\llbracket {{\mathbf{x}}_k|y_{0:k}}\rrbracket$ at ${k \in {\mathbb{N}_0}}$.
	\end{remark}
	%
	
	% \begin{definition}[Inclusion Property]\label{def_inlcusiondo}
	% 	$\forall i \in V$, a DSMF satisfies the inclusion property, if $\mathcal{B}_i({{\mathbf{x}}_k} 
	% 	) \supseteq\llbracket {{\mathbf{x}}_k|y_{0:k}} 
	% 	\rrbracket$ for $k \in {\mathbb{N}_0}$.
	% \end{definition}

	% In this work, we consider a commonly used assumption as follows.

	% \begin{assumption}[\!\!\cite{9094233},\cite{8552364}]\label{ass_inculsion}
	% 	$ \forall i \in V$, the initial prior belief of sensor $i$ (at $k=0$) satisfies
 %        %
 %        \[{\mathcal{B}_i^-({{\mathbf{x}}_0} 
	% 		)}  \supseteq \llbracket {{\mathbf{x}}_0} 
	% 	\rrbracket,\]
 %        where $\llbracket {{\mathbf{x}}_0} 
	% 	\rrbracket$ describes all the possible initial states $x_0$.
	% \end{assumption}
	% %
	% \begin{remark}
	% 	\aspref{ass_inculsion} can avoid resulting empty set (see \cite{cong2022stability} for more details). 
	% 	%
	% 	We can choose a large enough belief to satisfy \aspref{ass_inculsion} or use the techniques in \cite{cong2022stability} to deal with incorrect initial guesses, which is beyond the scope of this paper.
	% 	%
	% 	Unless otherwise stated, the results and discussions in the rest of this paper are under \aspref{ass_inculsion}.
	% \end{remark}
        %However, satisfying only the inclusion property for a DSMF is insufficient.
        
        However, the belief of sensor $i$ from DSMF may go unbounded as time elapses.
	Thus, for the synthesis of DSMFs, it is necessary to consider the asymptotic boundedness of the belief, as described in the following definition.
	\begin{definition}[Asymptotic Boundedness of DSMFing]\label{def_boundeddo}
		A DSMF is asymptotically bounded if, $\forall i \in V$
		\begin{equation}\label{eq_boundedness_definition}
			\mathop {\overline {\lim } }\limits_{k \to \infty } D(\mathcal{B}_i({{\mathbf{x}}_k} 
			)) < \infty.
		\end{equation}
	\end{definition}	 
	%
	% Due to the conservativeness of the framework, the present literature cannot provide a boundedness analysis or a boundedness condition for DSMFing frameworks.

        % As a key issue in boundedness analysis, the boundedness condition of DSMFing has not been investigated.
        % %
	In this work, we focus on the asymptotic boundedness condition of the DSMFing defined in \defref{def_boundeddo}.

	\section{Asymptotic Boundedness Condition for Distributed Set-Membership Filtering}\label{sec:sufficiency condition}
    
	To investigate the asymptotic boundedness condition, a general framework of classical DSMFing is established in \secref{sec_smfodoframework}.
        In \secref{sec_COIT}, we introduce a new concept -- Collective Observation-Information Tower (COIT), which incorporates commonly known information of measurements (in each ``group''). 
        Based on the COIT, an easy-to-check asymptotic boundedness condition of classical DSMFing is presented in \secref{sec_smfodoexist}.

	\subsection{Classical Distributed Set-Membership Filtering}\label{sec_smfodoframework}
	
        We establish a general framework of classical DSMFing using uncertain variables.
	First, we classify the sensors in $G$ that have a directed path to $i$ into sets of $\rho$-hop predecessors defined as follows.
	
	\begin{definition}[Set of $\rho$-Hop Predecessors]\label{def_hopneighbours}
            For a sensor $i$ in graph $G$, the set of all sensors having a direct path to $i$ involving $\rho$ edges is the set of $\rho$-hop predecessors of sensor $i$, labeled as $\mathcal{N}_{i,\rho}$.
            We set $\mathcal{N}_{i,\rho} = \emptyset$ for $\rho > \bar{\rho}_i$, where $\bar{\rho}_i$ stands for the eccentricity\footnote{The eccentricity of sensor $i$ is the maximum graph distance between $i$ and any other sensors in $G$.} of sensor $i$.
	\end{definition}
	Let ${{\mathcal{N}_{i,0}}} := \{i\}$, and we define $\mathcal{M}_j^i := \bigcup\nolimits_{\rho = 0}^{j + 1} {{\mathcal{N}_{i,\rho}}}$.
    Thus, $\mathcal{M}_{\bar\rho _i-1}^i$ includes sensor $i$ and all sensors that have a directed path to sensor $i$.
	Recall that sensor $i$ can only receive information from its neighbors, i.e., ${\mathcal{N}_{i,1}}$.
    Through this, a general framework for classical DSMFing for sensor $i$ is presented in \algref{alg:thmframe}.
    %, followed by a detailed line-by-line explanation.

        \begin{algorithm}
		\begin{footnotesize}
				\caption{Classical Distributed SMFing of Sensor $i \in V$}\label{alg:thmframe}
				\begin{algorithmic}[1]
						\STATE  \textbf{Initialization:}
						Initial prior belief $\mathcal{B}_i^-({{\mathbf{x}}_{0}})$.\label{line:thmframe - initialization}
						% \LOOP
                            \IF {$k > 0$}
                            \STATE \textbf{Local prediction step:}
                            \begin{equation}\label{eq_distrframe1}
                                {\mathcal{B}_i^-}({{\mathbf{x}}_{k + 1}} 
                    		)= A\mathcal{B}_i( {{\mathbf{x}}_k} 
                    		) + B\llbracket {{{\mathbf{w}}_k}} 
                    		\rrbracket;
                            \end{equation}\label{line_pre}
						\ENDIF\label{line_pre2}
                            \STATE \textbf{Local update step:} \begin{equation}\label{eq_distrframe2}
                                {\mathcal{B}_i^+}( {{\mathbf{x}}_{k}} )= {\mathcal{X}_{k}}({C_i},y_{k}^i,\llbracket {{\mathbf{v}}_{k}^i} \rrbracket) \cap {\mathcal{B}_i^-}( {{\mathbf{x}}_{k}} ),
                            \end{equation}
                            where
                            \begin{equation}\label{eq_ker_dsf}
                    		\begin{split}
                    			{\mathcal{X}_k}({C_i},y_{k}^i,\llbracket {{\mathbf{v}}_k^i} 
                    			\rrbracket) &= \{ x_k:y_{k}^i = {C_i}x_k + v_k^i,v_k^i \in \llbracket {{\mathbf{v}}_k^i} 
                    			\rrbracket\}
                    			\\&= \ker ({C_i}) + C_i^\dag  (\{ y_{k}^i\}  + \llbracket { - {\mathbf{v}}_k^i} 
                    			\rrbracket);	
                    		\end{split}
                    	\end{equation}\label{line_up}
						\STATE \textbf{Fusion step:}
                            \begin{equation}\label{eq_distrframe3}
                                \mathcal{B}_i({{\mathbf{x}}_{k}})= \bigcap\limits_{j\in{\mathcal{M}_{0}^i}} {{\mathcal{B}_j^+}( {{\mathbf{x}}_{k}})}.
                            \end{equation}\label{line_pro}
						% \ENDLOOP
					\end{algorithmic}
			\end{footnotesize}
	\end{algorithm}

        A detailed line-by-line description of \algref{alg:thmframe} is given as follows.
        Line~\ref{line:thmframe - initialization} initializes the algorithm.
        When $k>0$, Lines~\ref{line_pre} and~\ref{line_pre2} provide the prior belief $\mathcal{B}_i^-({{\mathbf{x}}_{k}})$ of the DSMF of sensor $i$, which is called the \textbf{Local prediction step}. 
        Line~\ref{line_up}, known as the \textbf{Local update step}, calculates the local posterior belief $\mathcal{B}_i^+( {{\mathbf{x}}_{k}})$ based on the local measurement $y_k^i$.
	Note that equation \eqref{eq_distrframe2} degenerates to ${\mathcal{B}_i^+}({{\mathbf{x}}_{k}} 
	)={\mathcal{B}_i^-}({{\mathbf{x}}_{k}} 
	)$ if no measurements are taken by $i$. 
	Line~\ref{line_pro} defines the \textbf{Fusion step}, where the local posterior beliefs at time $k$ that received from neighbors ${\mathcal{N}_{i,1}}$ are fused with the local posterior belief of sensor $i$, and returns the fused posterior belief $\mathcal{B}_i( {{\mathbf{x}}_{k}})$.	
    Most existing DSMFing methods adhere to the classical DSMFing framework outlined in \algref{alg:thmframe}, although they differ in the strategy of communication protocol, such as event-trigger\cite{8114330}, reduced-order \cite{8605375,ierardi2021distributed}, and coding–decoding \cite{Liu2021}.
        In what follows, the asymptotic boundedness of DSMFing is investigated under the framework of \algref{alg:thmframe}.

        \subsection{Collective Observation-Information Tower}\label{sec_COIT}

        Let $G_1^{\mathrm{s}}, \dots, G_T^{\mathrm{s}}$ be the source components\footnote{Given a directed graph $G = (V, E)$, a source component $G_t^{\mathrm{s}} =(V_t^{\mathrm{s}}, E_t^{\mathrm{s}})$ is defined as a strongly connected component of $G$ such that there are no edges from $V \backslash V_t^{\mathrm{s}}$ to $V_t^{\mathrm{s}}$\cite{7463019}.} of $G$.
        In this subsection, we introduce the concept of the Collective Observation-Information Tower (COIT) of the source component, which enables the analysis of the asymptotic boundedness of DSMFing (see \secref{sec_smfodoexist}).
        %
        % For DSMFing, we know that the asymptotic boundedness of all sensors in a directed graph is affected by the sensors in the source components.
        %
        % Assume there are $T$ source components $G_1^{\mathrm{s}}, \dots, G_T^{\mathrm{s}}$ of $G$.
        % %

        To start with, we propose an intersection-based outer bound of $\mathcal{B}_i({\mathbf{x}}_k)$ for sensor $i$ in source component $G_t^{\mathrm{s}}$ as follows.
        %
 %    	According to \algref{alg:thmframe}, the asymptotic boundedness of the fused posterior belief $\mathcal{B}_i( {{\mathbf{x}}_{k}})$  is determined by the measurements sequence $y_k^i$ and initial prior belief ${\mathcal{B}_i^-({{\mathbf{x}}_0})}$ w.r.t. $i\in{\mathcal{\bar N}_i}$.
 %        %
	% Through that, we can derive an outer bound of $\mathcal{B}_i( {{\mathbf{x}}_{k}} )$ in the following proposition.

\begin{figure}[h]
    		\centering
    		\includegraphics[width=1\linewidth]{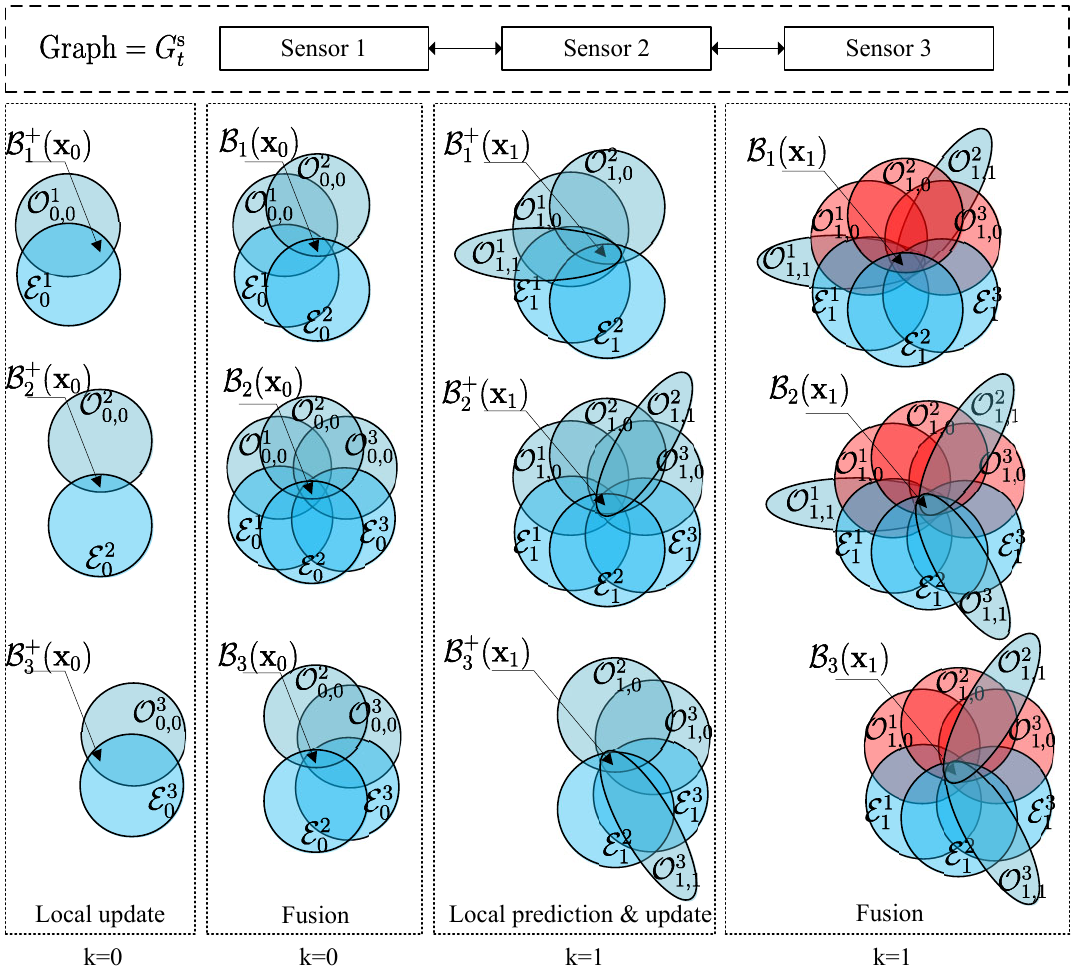}
     		\caption{Illustration of \propref{prop_COITbound} and the COIT for a source component $G_t^{\mathrm{s}}$.
            % The geometric interpretation of the believes in \algref{alg:thmframe} and their outerbounds, considering a simple graph $G_t^{\mathrm{s}}$:
            %
            At $k=0$, each sensor obtains a local posterior belief $\mathcal{B}_i^+( {{\mathbf{x}}_{0}})$ through the local update step \eqref{eq_distrframe2} in \algref{alg:thmframe}, which is exactly $\mathcal{O}_{0,0}^i \bigcap \mathcal{E}_0^i$. 
            After the fusion step defined by~\eqref{eq_distrframe3}, the fused posterior belief $\mathcal{B}_i({{\mathbf{x}}_{0}})$ is outer bounded by $\big({\bigcap_{l \in \mathcal{M}_0^i} {\mathcal{O}_{0,0}^l} }\big)  \bigcap \big(\bigcap_{l \in \mathcal{M}_0^i} {\mathcal{E}_0^l}\big)$, which satisfies~\eqref{eq_prop_COITbound}.
            %each sensor combines its local posterior belief with the local posterior beliefs received from its neighbors to obtain the fused posterior belief $\mathcal{B}_i({{\mathbf{x}}_{0}})$. 
            %
            At $k=1$, through \eqref{eq_distrframe1} and \eqref{eq_distrframe2}, each sensor predicts and update the local posterior belief $\mathcal{B}_i^+( {{\mathbf{x}}_{1}})$.
            Then, through the fusion step~\eqref{eq_distrframe3}, the fused posterior belief $\mathcal{B}_i({{\mathbf{x}}_{1}})$ is outer bounded by $\big({\bigcap_{r=0}^1\bigcap_{l \in \mathcal{M}_{1-r}^i} {\mathcal{O}_{1,r}^l} }\big)  \bigcap \big(\bigcap_{l \in \mathcal{M}_1^i} {\mathcal{E}_1^l}\big)$, where the COIT is $\bigcap \nolimits_{i \in V_t^{\mathrm{s}}} {\cal O}_{1,0}^i$ (marked in red).
            The COIT contains the commonly known observation information sets in a collective manner.
            As a counterexample, even though $\mathcal{O}_{1,1}^2$ is also commonly known to the three sensors (as $\tilde{\rho}_2 = 1$), $\mathcal{O}_{1,1}^1$ and $\mathcal{O}_{1,1}^3$ are not (since $\tilde{\rho} =2$, $\tilde{\rho}_1=2$, $ \tilde{\rho}_3 = 2$);
            based on~\eqref{eq_coit}, at $k = 1$, the COIT can at most include the observation information (from the collection of these three sensors) up to time $k - \tilde{\rho} + 1 = 0$.
            }
    		\label{f1}
    	\end{figure}

	\begin{proposition}[Intersection-Based Outer Bound]\label{prop_COITbound}
		Consider a source component $G_t^{\mathrm{s}} =(V_t^{\mathrm{s}}, E_t^{\mathrm{s}})$ in $G$.
        $\forall k \in {\mathbb{N}_0}$, an outer bound of $\mathcal{B}_i( {{\mathbf{x}}_{k}})$ ($i\in V_t^{\mathrm{s}}$) derived by \algref{alg:thmframe} is
            \begin{equation}\label{eq_prop_COITbound}
                \mathcal{B}_i({\mathbf{x}}_k) \subseteq
                \Bigg(\bigcap\limits_{r = 0}^k {\bigcap\limits_{l \in \mathcal{M}_{k - r}^i}\!\!{\mathcal{O}_{k,r}^l} }\Bigg)  \bigcap \Bigg(\bigcap_{l \in \mathcal{M}_{k}^i} {\mathcal{E}_k^l}\Bigg),
            \end{equation}
		where $\mathcal{M}_{j}^i=V_t^{\mathrm{s}}$ when $j\geq\bar\rho_i-1$.        
		% \begin{equation}\label{eq_def_OIT} 
		% 		\mathcal{C}_k^{i} =\bigcap\limits_{r = 0}^{k - {\bar\rho }+1} {\bigcap\limits_{l\in V_t^{\mathrm{s}}} { { {\mathcal{O}_{k,r}^l}} } }\bigcap\bigcap\limits_{r =k-{\bar\rho}+2}^{k-{\bar\rho_i}+1} {\bigcap\limits_{l\in V_t^{\mathrm{s}}} { { {\mathcal{O}_{k,r}^l}} } },			
		% \end{equation}
  %       %
  %       \begin{equation} \label{eq_qbarOIT}      
  %           {\mathcal{Q}}_{k,k'}^i = {\bigcap\limits_{j = k'}^{k} {\mathcal{O}_{k,j}^i} \cap \mathcal{E}_k^i}\bigcap\bigcap\limits_{r = 0}^{k-k'} {\bigcap\limits_{l\in{\mathcal{N}_{i,r+1}}} {\left( { {\bigcap\limits_{j = k'}^{k - r} {\mathcal{O}_{k,j}^l} } \bigcap \mathcal{E}_k^l} \right)} },
  %       \end{equation}
  %      and $\bar\rho\geqslant\bar\rho_i\geqslant2$ denotes the diameter of $G_t^{\mathrm{s}}$.
        In~\eqref{eq_prop_COITbound},
		\begin{equation}\label{eq_observationinformationtdistri1}
			{{\mathcal{O}}_{k,r}^l}: =  {A^{k - r}}{\mathcal{X}_r}(C_l,{y_r^l},\llbracket {{{\mathbf{v}}_r^l}} 
			\rrbracket) + \!\!\sum\limits_{\tau = r}^{k - 1} \!\!{{ A^{k - 1 - \tau}}B\llbracket {{{\mathbf{w}}_\tau}} 
				\rrbracket}
		\end{equation}
            is the observation-information set~\cite{cong2022stability} at time $k \geqslant r$ contributed by $y_r^l$;
		\begin{equation}\label{eq_observationinformationtdistri2}
			{\mathcal{E}}{_{k}^l}:={ A^{k}}{\mathcal{B}_l^-({{\mathbf{x}}_0} 
				)} + \sum\limits_{\tau = 0}^{k- 1} {{A^{k - 1 - \tau}}B\llbracket {{{\mathbf{w}}_\tau}} 
				\rrbracket}
		\end{equation}
            is the state-evolution set~\cite{cong2022stability} at time $k$. 
	\end{proposition}
	\begin{Proof}
		See \apxref{apx_proCOITbound}.
	\end{Proof}
        The geometric interpretation of \propref{prop_COITbound} is shown in \figref{f1}.
        At time $k$, sensor $i$ can indirectly obtain the observation information up to $k-\tilde{\rho}_i+1$ (where $\tilde{\rho}_i = \max\{\bar{\rho}_i, 1\}$), i.e., $$k - \bar{\rho}_i + 1~\textnormal{for}~|V_t^{\mathrm{s}}| > 1~\textnormal{and}~k - \bar{\rho}_i~\textnormal{for}~|V_t^{\mathrm{s}}| = 1.$$
        This fact defines the COIT of $G_t^{\mathrm{s}}$ as follows
        %
        % By defining $\bar\rho$ as the diameter of $G_t^{\mathrm{s}}$, for all sensors in $G_t^{\mathrm{s}}$, the public part of these observation information can be defined as
        \begin{equation}\label{eq_coit}
            \mathcal{C}_k^{(t)}:=\bigcap\limits_{r = 0}^{k-\tilde{\rho}+1} {\bigcap\limits_{l \in V_t^{\mathrm{s}}} {\mathcal{O}_{k,r}^l}},
        \end{equation}
        where $\tilde{\rho} =  \max\{\tilde{\rho}_i\colon i \in V_t^{\mathrm{s}}\}$, which includes the commonly known observation-information sets in a collective/complete manner.
        An example of the COIT can be found in \figref{f1} (see the red circles).
        Note that $\forall i\in V_t^{\mathrm{s}}$, we have $\mathcal{B}_i({\mathbf{x}}_k) \subseteq \mathcal{C}_k^{(t)}$.
        
        % which is the COIT of $G_t^{\mathrm{s}}$ (marked by red in \figref{f1}).
 %        In the filtering process, after $\bar\rho_i$ steps, sensor $i$ can receive all measurement information at $k-\bar\rho_i$ from the sensor network, which means the filtering information at $k-\bar\rho_i$ has achieved consensus for sensor $i$.
 %        %
 %        All filtering information that already achieved consensus is denoted as $\mathcal{Q}_k^i$ for $k \geqslant {\bar\rho _i}$, by \propref{prop_COITbound}.
 %        %
 %        Note that 
	% \begin{equation}\label{eq_14}
	% 	\mathcal{Q}_k^i = \bigcap\limits_{r = 0}^{k - {\bar\rho _i}} {\bigcap\limits_{l\in\mathcal{\bar N}_i} { { {\mathcal{O}_{k,r}^l} } } }   
	% 	\subseteq \bigcap\limits_{r = k - {\bar\rho _i} - \delta }^{k - {\bar\rho _i}} {\bigcap\limits_{l\in\mathcal{\bar N}_i} {\mathcal{O}_{k,r}^l} }
	% \end{equation}
	% holds for any $\delta  \in [0,k]$. Here, we define the right-hand side of \eqref{eq_14} as the COIT in the following definition.
	% %
	% \begin{definition}[Distributed Observation-Information Tower]\label{def_COIT}
	% 	The COIT at time $k \in [{\bar\rho _i}, + \infty )$ for sensor $i$ is defined by the intersection of the observation-information sets as:
	% 	\begin{equation}\label{eq_def_OIT}
	% 		\mathcal{C}_k^{i}= \bigcap\limits_{r = 0}^{k - {\bar\rho _i}} {\bigcap\limits_{l\in\mathcal{\bar N}_i}{\mathcal{O}_{k,r}^l} }. 
	% 	\end{equation} 
	% \end{definition}
	% %
	%

	\subsection{Asymptotic Boundedness of Classical DSMFing}\label{sec_smfodoexist}

    For the source component ${G_t^{\mathrm{s}}}$, let $i_{(t,m)}$ denote the $m$th sensor in $t$, we lump the measurement equations of all sensors in ${V_t^{\mathrm{s}}} = \{ i_{(t,1)}, \ldots ,i_{(t,m)}\}$ together, i.e.,
    \begin{align}\label{eq_intro_lumpmeaurement}
        {{\mathbf{y}}_{k}^{(t)}}=&\ \textnormal{col}( {{\mathbf{y}}_k^{i_{(t,1)}}},\dots,  {{\mathbf{y}}_k^{i_{(t,m)}}} )\notag\\%{\begin{bmatrix}
            %    {{\mathbf{y}}_k^{i_{t,1}}} \\ 
            %    \vdots  \\ 
            %    {{\mathbf{y}}_k^{i_{t,m}}} 
       % \end{bmatrix}} 
       =& \underbrace{ {\begin{bmatrix}
                {{C_{i_{(t,1)}}}} \\ 
                \vdots  \\ 
                {{C_{{i_{(t,m)}}}}} 
        \end{bmatrix}} {{\mathbf{x}}_k} + {\begin{bmatrix}
                {{\mathbf{v}}_k^{i_{(t,1)}}} \\ 
                \vdots  \\ 
                {{\mathbf{v}}_k^{{i_{(t,m)}}}} 
        \end{bmatrix}} }_{ C^{(t)}{{\mathbf{x}}_k} + {{\mathbf{v}}_{k}^{(t)}}},
    \end{align}
    where $C^{(t)}$ is the joint measurement matrix w.r.t. $G_t^{\mathrm{s}}$.
Then, we perform an observability decomposition for \eqref{eq_intro_system} and \eqref{eq_intro_lumpmeaurement}, i.e., there exists a nonsingular ${P^{(t)}} = \textnormal{col}({{{P_o^{(t)} }}, {P_{\bar o}^{(t)} })}\in \mathbb{R}^{n\times n}$ such that the equivalence transformation   \begin{align*}\tilde{\mathbf{x}}_k &=\textnormal{col}(\tilde{\mathbf{x}}_k^o, \tilde{\mathbf{x}}_k^{\bar{o}})\\
&=\left[\begin{matrix}P_o^{(t)}\mathbf{x}_k\\
P_{\bar o}^{(t)}\mathbf{x}_k\end{matrix}\right] =\underbrace{\left[\begin{matrix}P_o^{(t)}\\
P_{\bar o}^{(t)}\end{matrix}\right]}_{ P^{(t)}} \mathbf{x}_k\end{align*}  transforms \eqref{eq_intro_system} and \eqref{eq_intro_lumpmeaurement} into
	\begin{subequations}
        \begin{align}
    		{\begin{bmatrix}
    					{{\tilde{\mathbf{x}}}_{k+1}^{o}}\\ 
    					{{\tilde{\mathbf{x}}}_{k+1}^{\bar o}} 
    			\end{bmatrix}} &= {\begin{bmatrix}
    					{{{\tilde A}_{o}^{(t)}}}&0 \\ 
    					{{{\tilde A}_{21}^{(t)}}}&{{{\tilde A}_{\bar o}^{(t)}}} 
    			\end{bmatrix}}{\begin{bmatrix}
    					{{\tilde{\mathbf{x}}}_k^{o}}\\ 
    					{{\tilde{\mathbf{x}}}_k^{\bar o}} 
    			\end{bmatrix}} + {\begin{bmatrix}
    					{{{\tilde B}_{o}^{(t)}}}\\ 
    					{{{\tilde B}_{\bar o}^{(t)}}} 
    			\end{bmatrix}} {{\mathbf{w}}_k},\label{eq_ob_sys}\\ 
    		{{\mathbf{y}}_k^{(t)}} &= {\begin{bmatrix}
    					{{{\tilde C_{o}^{(t)}}}}&0 
    			\end{bmatrix}}{{{\tilde{\mathbf{x}}}}_k}+ {{\mathbf{v}}_{k}^{(t)}}, \label{eq_ob_sysmeasure}
    	\end{align}
    \end{subequations}
	%
	% where 
	% \begin{equation}\label{transform_matrix}			
	% 		\tilde C^{(t)} = C^{(t)} {(P^{(t)})^{-1}}=  {\begin{bmatrix}
	% 				{{{\tilde C_{o}^{(t)}}}}&0 
	% 		\end{bmatrix}};
	% \end{equation}
	% %
    where $\tilde{\mathbf{x}}_k^o \in {\mathbb{R}^{n_o}}$, $\tilde{\mathbf{x}}_k^{\bar o} \in {\mathbb{R}^{n_{\bar o}}}$, and $({{{\tilde A}_{o}^{(t)}}}, {\tilde C_{o}^{(t)}})$ is observable.
	%	
	%
	% Then, based on \eqref{eq_ob_sys} and \eqref{eq_ob_sysmeasure}, the DSMFing framework \eqref{eq_distrframe1}-\eqref{eq_distrframe3} for sensor $i\in G_t^{\mathrm{s}}$ can be transformed into 
	% \begin{align}	
	% 	{\mathcal{B}_i^-}({{\tilde{\mathbf{x}}}_{k + 1}} 
	% 	) &=  \tilde A^{(t)}{\mathcal{B}_i}({{\tilde{\mathbf{x}}}_{k}} 
	% 	) + \tilde B^{(t)}\llbracket {{{\mathbf{w}}_k}} 
	% 	\rrbracket,\label{eq_distrframedec1}\\	
	% 	{\mathcal{B}_i^+}({{\tilde{\mathbf{x}}}_{k}} 
	% 	) &= {\mathcal{\tilde X}_k^o}({{\tilde C}_i},y_{k}^i,\llbracket{\mathbf{v}}_k^i \rrbracket) \cap {\mathcal{B}_i^-}({{\tilde{\mathbf{x}}}_{k}} 
	% 	),\label{eq_distrframedec2}\\	
	% 	{\mathcal{B}_i}({{\tilde{\mathbf{x}}}_{k}} 
	% 	) &=\bigcap\limits_{j \in\mathcal{M}_{0}^i} {{\mathcal{B}}_j^+({{\tilde{\mathbf{x}}}_{k}} 
	% 		)},\label{eq_distrframedec3}	
	% \end{align}
	% where ; \[\tilde {\mathcal{X}}_k^o({{\tilde C}_{i}},{y_k},\llbracket {{{\mathbf{v}}_k^i}} 
 %    \rrbracket) = \ker (\tilde C_{i}) + {\tilde C_{i}^\dag }(\{ {y_k}\}  + \llbracket { - {{\mathbf{v}}_k^i}} 
 %    \rrbracket).\]	
%
Then, based on \propref{prop_COITbound}, an decomposition-based outer bound of $\mathcal{B}_i(\mathbf{x}_k)$ for $i\in V_t^{\mathrm{s}}$ is presented (see \propref{prop_decompCOIT}).

\begin{proposition}[Observability Decomposition-Based Bound]\label{prop_decompCOIT}
	When $k > \tilde{\rho}$, $\mathcal{B}_i({{\mathbf{x}}_{k}})$ for $i\in V_t^{\mathrm{s}}$ is outer bounded by
        \begin{equation}\label{eq_propdecopCOIT_outbo}
		\mathcal{B}_i(\mathbf{x}_k) \subseteq (P^{(t)})^{-1} \big(P_o^{(t)} \mathcal{C}_k^{(t)} \times \mathcal{\tilde S}_k^{i}\big),
	\end{equation}
	% \begin{equation}\label{eq_propdecopCOIT_outbo}
	% 	\mathcal{B}_i({{\tilde{\mathbf{x}}}_{k}}) \subseteq {\mathcal{\tilde C}_k^{(t)} \times \mathcal{\tilde S}_k^{i}},
	% \end{equation}
	where
    %$\mathcal{\tilde C}_k^{(t)} = P_o^{(t)} \mathcal{C}_k^{(t)}$ and
    % \begin{equation}\label{eq_coitobser}
    %         \mathcal{\tilde C}_k^{(t)}=\bigcap\limits_{r = 0}^{k-\bar\rho+1} {\bigcap\limits_{l \in V_t^{\mathrm{s}}} \mathcal{\tilde{O}}_{k, r}^{{l,o}}},
    % \end{equation}
    %
    \begin{multline}\label{eq_prop2_sij}
        \mathcal{\tilde S}_k^{i} = (\tilde A_{\bar o}^{(t)})^k P_{\bar{o}}^{(t)} \mathcal{B}_i^-(\mathbf{x}_0)  \\
        +\sum\limits_{j = 0}^{k - 1} {(\tilde A_{\bar o}^{(t)})^{k - 1 - j}\Big({\tilde A_{21}^{(t)}} P_{o}^{(t)} \mathcal{B}_i(\mathbf{x}_j) + {\tilde B_{\bar o}^{(t)}}\llbracket {{{\mathbf{w}}_j}}\rrbracket\Big)}.
    \end{multline}
\end{proposition}
\begin{Proof}
	See \apxref{apx_decompo_COITb}.
\end{Proof}
Through \propref{prop_decompCOIT}, we are ready to provide an asymptotic boundedness condition for the classical DSMFing.
\begin{theorem}[Asymptotic Boundedness of DSMF]\label{thm_boundednesssuffcient}
	% With a bounded initial uncertainty, i.e., $D(\mathcal{B}_i^-({{\mathbf{x}}_0})) < \infty$, 
    $\forall i \in V$ with $D(\mathcal{B}_i^-({{{\mathbf{x}}}_{0}}))<\infty$, $\mathcal{B}_i({{\mathbf{x}}_{k}} 
	)$ derived from \algref{alg:thmframe} satisfies~\eqref{eq_boundedness_definition},
    % %
    % $\mathop {\overline {\lim } }\nolimits_{k \to \infty } D(\mathcal{B}_i({\mathbf{x}}_{k}))<\infty$,
	if $\forall G_t^{\mathrm{s}} \subseteq G$, the following two conditions hold:
    %the system described by \eqref{eq_ob_sys} and \eqref{eq_ob_sysmeasure} satisfy: 		
	\begin{itemize}
		\item[(i)] ${{\tilde A}_{\bar o}^{(t)}}$ is marginally stable;
		\item[(ii)] For all eigenvalues $\lambda$ of ${{\tilde A}_{\bar o}^{(t)}}$ with $|{\lambda }| = 1$,
            \begin{equation}\label{eq_filteabilitycondi}
                \!\!\!\!\!\!\!\!\!\!\!\!\!\!{\rm rank}\!\left({\begin{bmatrix}
					{{{\tilde A}_{\bar o}^{(t)}} \!\!-\! \lambda I_{n_{\bar{o}}}}&\!{{{\tilde B}_{\bar o}^{(t)}}}&{{{\tilde A}_{21}^{(t)}}} 
			\end{bmatrix}}\right)
			\!=\! {\rm rank}\!\left(
					{{{\tilde A}_{\bar o}^{(t)}} \!\!-\! \lambda I_{n_{\bar{o}}}}\right)\!.
            \end{equation}
		% \begin{multline}\label{eq_f ilteabilitycondi}
		% 	{\rm rank}({\begin{bmatrix}
		% 			{{{\tilde A}_{\bar o}^{(\mathcal{A})}} - \lambda I}&{{{\tilde B}_{\bar o}^{(\mathcal{A})}}}&{{{\tilde A}_{21}^{(\mathcal{A})}}} 
		% 	\end{bmatrix}})
		% 	\\= {\rm rank}(
		% 			{{{\tilde A}_{\bar o}^{(\mathcal{A})}} - \lambda I}).
		% \end{multline}
	\end{itemize}
\end{theorem}
\begin{Proof}
	See \apxref{apx_thmoundednesssuffcient}.
\end{Proof}
Moreover, we have the following corollary.
\begin{corollary}\label{coro_1}
    $\forall G_t^{\mathrm{s}}$ of $G$, if the pair $(A,C^{(t)})$ is detectable, DSMFing can achieve asymptotic boundedness.
\end{corollary}
\begin{Proof}
    If the pair $(A,C^{(t)})$ is detectable.
    $\forall G_t^{\mathrm{s}}$ of $G$, it means that ${{{\tilde A}_{\bar o}^{(t)}}}$ is Schur stable and ${{{\tilde A}_{\bar o}^{(t)}} \!\!-\! \lambda I}$ has full row rank.
    Thus, conditions (i) and (ii) in \thmref{thm_boundednesssuffcient} hold, and \begin{align}        
    \mathop {\overline {\lim } }\nolimits_{k \to \infty } D(\mathcal{B}_i({\mathbf{x}}_{k}))<\infty, & &\forall i\in V.\end{align}
\end{Proof}
\begin{remark}
    The condition in \corref{coro_1} is the well-known \emph{collective detectability condition}, which guarantees the asymptotic convergence of DOs \cite{7463019} and the asymptotic boundedness of the error covariance of DKF \cite{BATTISTELLI2014707}.
    This condition is proven to be a special case of \thmref{thm_boundednesssuffcient}.
    It implies that for certain unobservable substates that are not Schur stable in \eqref{eq_intro_system}, deploying additional sensors is not always necessary, since bounded DSMFing can still be achieved as long as \thmref{thm_boundednesssuffcient} holds. This property is a unique advantage of DSMFing compared to DKFs.
\end{remark}

\section{Numerical Example}\label{sec_num}

\begin{figure}[ht]
	\centering
	\includegraphics[width=0.9\linewidth]{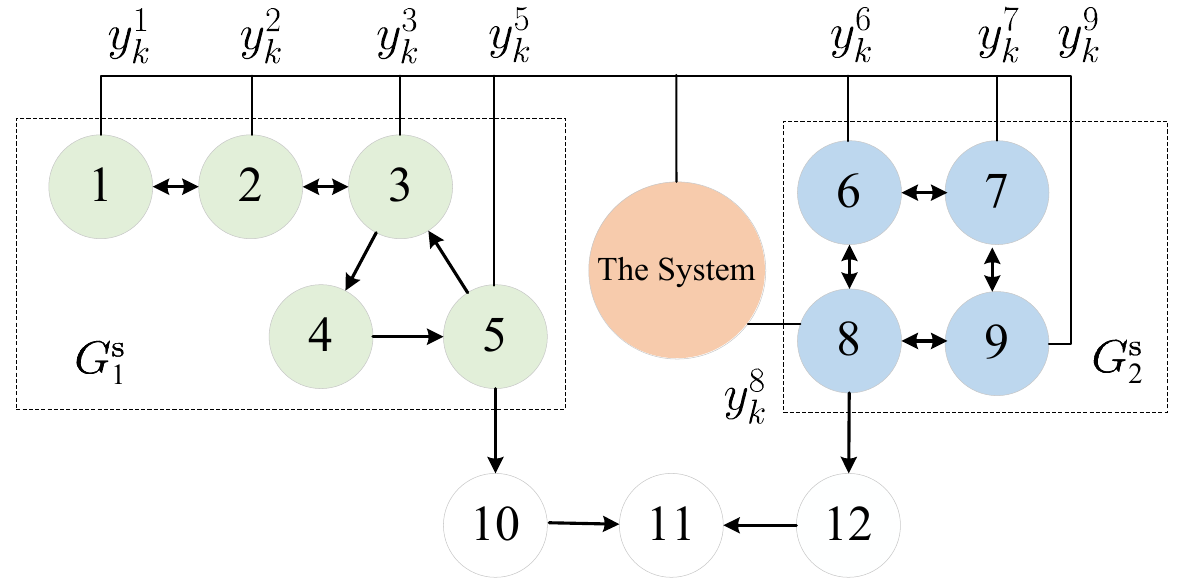}
	\caption{A system measured by a sensor network with $12$ sensors.
    Sensors $1$-$9$ directly obtain the measurements, which compose the source components $G_1^{\mathrm{s}}$ and $G_2^{\mathrm{s}}$ marked in green and blue, respectively.
    Sensors $10$-$12$ cannot take measurements directly from the system, and are not in any source components.
    % Sensors in source component $G_1^{\mathrm{s}}$ are marked with green, while sensors in source component $G_2^{\mathrm{s}}$ are marked with blue. Sensors that are not in any source components are white. Moreover, sensors $1,2,3,5,6,7,8,9$ can obtain a measurement from the system plant.
    %
    }
	\label{fig_graph}
\end{figure}

In this section, we validate \thmref{thm_boundednesssuffcient} through a simulation example by implementing the classical DSMF in \algref{alg:thmframe} with constrained zonotopes \cite{Scott2016}.
Consider the network of 12 sensors shown in \figref{fig_graph}.
The system parameters and the joint measurement matrices (w.r.t. the $2$ source components) are
\begin{align*}%\label{eq_systemsimu_detect}
		A &=  {\rm diag}(
				{0.99}, 
				{1.01}, 
				{0.98}, 
				1,
				{0.8}, 
				{0.9}),\hfill \\B &= {\begin{bmatrix}
                1&0&1&0&1&0\\
                0&1&0&0&1&1\\
		\end{bmatrix}^\top}, \hfill \\
		{C^{{(1)} }}& = \underbrace{{\begin{bmatrix}
				1&0&0&0&0&0 \\ \hdashline
				0&{0.86}&0&0&0&0 \\ \hdashline
				0&1&{1.01}&0&0&0 \\ \hdashline
                0&0&0&0&0&0 \\ \hdashline
				0&{0.6}&0&{0.2}&0&0 
		\end{bmatrix}}}_{\textnormal{col}(C_1,C_2,C_3, C_4, C_5)} \begin{array}{c}
   \\[-0.8em]
  C_1\\
   C_2 \\
   C_3\\
   C_4\\
   C_5
\end{array}, \hfill \\
		{C^{{(2)}}}& =\underbrace{{\begin{bmatrix}
				{0.95}&0&0&0&0&0 \\ \hdashline
				0&{1.05}&0&0&0&0 \\ \hdashline
				0&0&2&0&0&0 \\ \hdashline
				0&0&0&0&1&0 
		\end{bmatrix}}}_{\textnormal{col}(C_6,C_7,C_8, C_9)}\begin{array}{c}
   \\[-0.8em]
  C_6\\
   C_7 \\
   C_8\\
   C_9
\end{array}, \hfill \\ 
        \llbracket{{\mathbf{w}}_{k}}\rrbracket& = [-1,1]^2,\quad \llbracket{{\mathbf{v}}_{k}^i}\rrbracket = [-1,1]. 
\end{align*}
% We set the initial prior belief as $\mathcal{B}_i({{\mathbf{x}}_{k}}) =[-10,10]^6$ for each sensor.
%
The joint subsystem formed by $G_1^{\mathrm{s}}$ satisfies the collectively detectable condition, while the joint subsystem formed by  $G_2^{\mathrm{s}}$ is not collectively detectable in the sense of \corref{coro_1}. However, it satisfies the conditions in \thmref{thm_boundednesssuffcient}.

\begin{figure}[ht]
	\centering
	\includegraphics[width=0.9\linewidth]{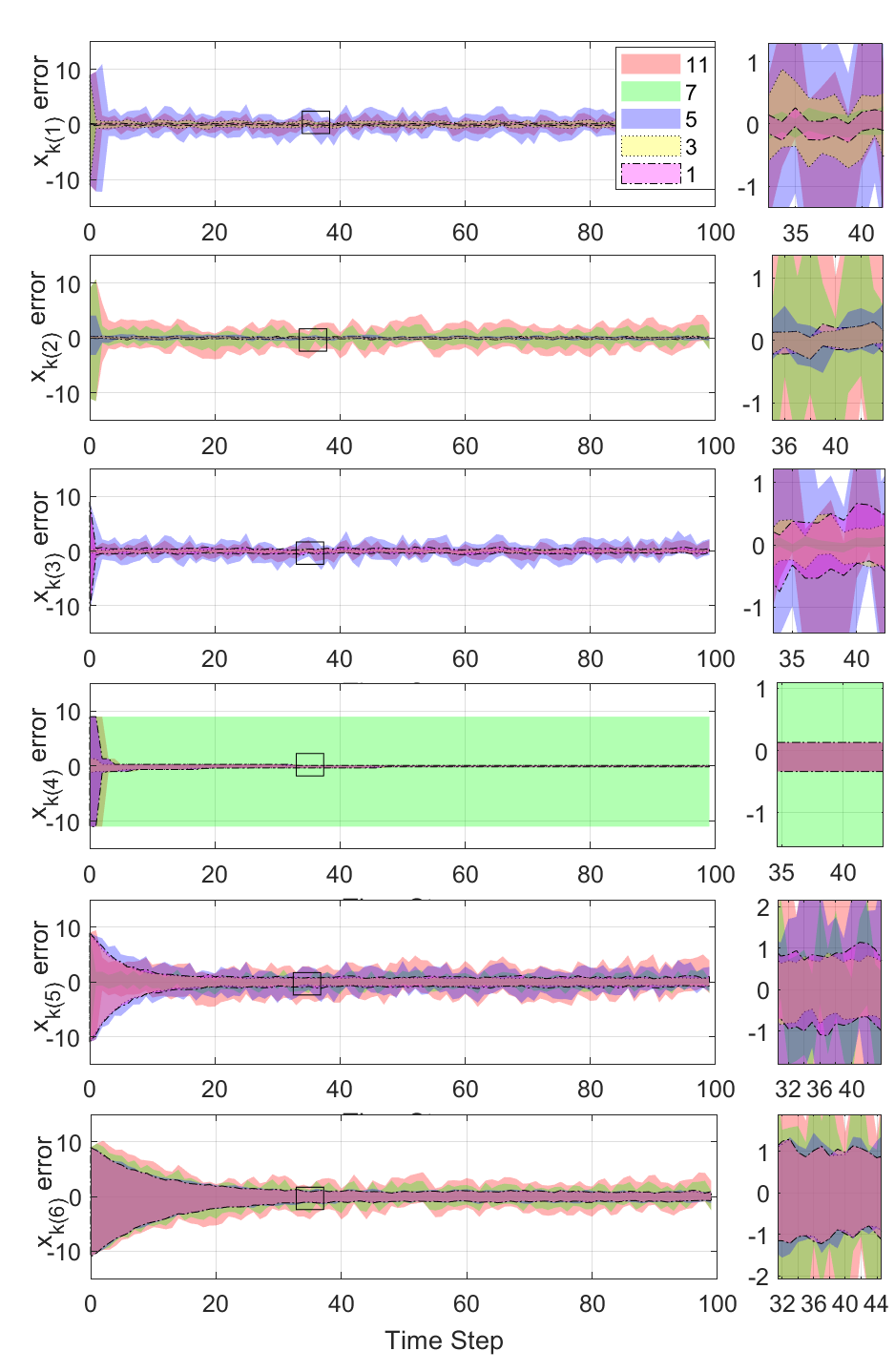}
	\caption{Estimation error range of each sensor for different components of system's state. From top to bottom, the estimation error bound of each component $x_{k(1)}$ to $x_{k(6)}$ for sensors ${1,3,5,7,11}$ are shown in each sub-figure, respectively. In each subfigure, the estimation error range of different sensors is shown using different colors, where the error range of each dimension for sensor $i$ is calculated by ${{\overline{\text{IH}}}}(\mathcal{B}_i({{\mathbf{x}}_{k(j)}}))-x_{k(j)}$ for $j\in[1,6]$. The details of each sub-figure are shown on the right.} 
	\label{fig_detect}
\end{figure}
Without loss of generality, we investigate the asymptotic boundedness of DSMF for sensors 1, 3, 5, 7, and 11, which are taken from different components of the system's state (see \figref{fig_detect}). 
Let $x_{k(i)}$ denote the $i$th component of $x_k$.
For sensors 1,3, and 5 in $G_1^{\mathrm{s}}$, we can see that sensor 5 has the worst estimation performance compared to all the observable dimensions, as it is located at the edge of the network.

% $x_{k(5)}$ and $x_{k(6)}$, we can see all the sensors in the network share similar estimation error bounds. 
% %
% For the observable dimensions, we know sensor 5 has the worst estimation performance compared to all the observable dimensions as it is located at the edge of the network.
%
For sensor 7 in $G_2^{\mathrm{s}}$, we can see that it can still achieve a bounded estimation error range for the marginally stable unobservable dimension $x_{k(4)}$.
It will not go unbounded because the disturbance does not influence the corresponding system channel, which verifies the sufficiency of the conditions in \thmref{thm_boundednesssuffcient}.

\section{Conclusion}\label{sce_conclusion}
In this paper, the asymptotic boundedness of DSMFing for linear discrete-time systems has been investigated. 
A novel concept, termed the COIT, has been introduced to characterize the fundamental relationship between the source components and the set estimates. 
Based on COIT, a sufficient condition for the asymptotic boundedness of general DSMFing has been established and shown to be weaker than the classical collective detectability condition for the convergence of DOs and bounded error covariance of DKFs, providing a guideline for the practical application of DSMFing. 
\appendices

\section{Proof of \propref{prop_COITbound}}\label{apx_proCOITbound}

Initially, we provide a lemma to support the following proof.

\begin{lemma}[\!\!\cite{cong2022stability}]\label{lem_set_intersect}
	For sets $\mathcal{S}_1,\cdots,\mathcal{S}_J$, and $\mathcal{T}$, we have
	\begin{equation}\label{eq_lm_minkintersect}
		\left( {\bigcap\limits_{j = 1}^J {{\mathcal{S}_j}} } \right) + \mathcal{T} \subseteq \bigcap\limits_{j = 1}^J {\left( {{\mathcal{S}_j} + \mathcal{T}} \right)}.
	\end{equation}
\end{lemma}

Then, we prove \eqref{eq_prop_COITbound} by mathematical induction.

\emph{Base case:}
When $k=0$, according to \eqref{eq_distrframe1}-\eqref{eq_distrframe3}, we have
\begin{align*}
    \mathcal{B}_i({\mathbf{x}}_0) &= \!\!\bigcap\limits_{l\in\mathcal{M}_{0}^i} \!\!{\left( {{\mathcal{X}_0}({C_l},y_0^l,\llbracket {{\mathbf{v}}_0^l}\rrbracket) \cap {\mathcal{B}_l^ - }({\mathbf{x}}_0)} \right)} \\
    &\stackrel{\eqref{eq_observationinformationtdistri1}}{=} \!\!\!\!\bigcap\limits_{l\in\mathcal{M}_{0}^i} \!\!{\left( {\mathcal{O}_{0,0}^l \cap \mathcal{E}_0^l} \right)},
\end{align*}
% \begin{equation}
% 	\begin{split}
% 		&\mathcal{B}_i({\mathbf{x}}_0) = \bigcap\limits_{l\in\mathcal{M}_{0}^i} {\left( {{\mathcal{X}_0}({C_l},y_0^l,\llbracket {{\mathbf{v}}_0^l} 
% 				\rrbracket) \cap {\mathcal{B}_l^ - }({\mathbf{x}}_0)} \right)}  \hfill \\
% 		% &= \bigcap\limits_{l\in\mathcal{M}_{0}^i} {\left( {{\mathcal{X}_0}({C_l},y_0^l,\llbracket {{\mathbf{v}}_0^l} 
% 		% 		\rrbracket)} \right)} \bigcap\bigcap\limits_{l\in\mathcal{M}_{0}^i} {\left( {{\mathcal{B}_l^ - }({\mathbf{x}}_0)} \right)}  \hfill \\
% 		&= \bigcap\limits_{l\in\mathcal{M}_{0}^i} {\left( {\mathcal{O}_{0,0}^l \cap \mathcal{E}_0^l} \right)}, \\ 
% 	\end{split}
% \end{equation}
which means \eqref{eq_prop_COITbound} holds for $k=0$.

\emph{Inductive step:}
Assume \eqref{eq_prop_COITbound} holds for $k=k'$, when $k=k'+1$, we have
\begin{equation}\label{eq_ap2_outb_1}
	\begin{split}
		\mathcal{B}_i({\mathbf{x}}_{k'+1}) 
        &= \bigcap\limits_{l\in\mathcal{M}_{0}^i} \left({ {{\mathcal{X}_{k'+1}}}({C_l},y_{k'+1}^l,\llbracket {{\mathbf{v}}_{k'+1}^l} 
				\rrbracket) \cap {\mathcal{B}_l^ - }({\mathbf{x}}_{k'+1})} \right)  \hfill \\
		&= \bigcap\limits_{l\in\mathcal{M}_{0}^i} \left({\mathcal{O}_{{k'+1},{k'+1}}^l \bigcap \left( {A{{\mathcal{B}_l }({\mathbf{x}}_{k'})}  + B\llbracket {{{\mathbf{w}}_{k'}}} 
				\rrbracket} \right)}\right)  \hfill \\
	  &\mathop \subseteq \limits^{\eqref{eq_appproof_1}} \left(\bigcap\limits_{r = 0}^{k'+1} {\bigcap\limits_{l \in \mathcal{M}_{k'+1 - r}^i}\!\!\!\!\!\!\!\!{\mathcal{O}_{k'+1,r}^l}}\right)  \bigcap \left(\bigcap\limits_{l \in \mathcal{M}_{k'+1}^i}\!\!\!\!{\mathcal{E}_{k'+1}^l}\right).  
	\end{split}
\end{equation}
Specifically, in the gray-shaded formula in \eqref{eq_appproof_1}, $(a)$ holds since $\bigcup\nolimits_{l \in \mathcal{M}_0^i} {\mathcal{M}_j^l = \mathcal{M}_{j + 1}^i}$ holds for $j<\bar\rho_i-1$.
%
% When $k\geqslant\bar\rho_i-1$, we use mathematical induction to prove \eqref{eq_prop_COITbound}.
% %

% \emph{Base case:}
% When $k=\bar\rho_i-1$, we have

% \begin{equation}\label{eq_ap2_outb_2}
% 	\begin{gathered}
% 		\mathcal{B}_i({\mathbf{x}}_{{\bar\rho _i}-1}) = \bigcap\limits_{l\in\mathcal{M}_{0}^i} {\left( {{\mathcal{X}_{{\bar\rho _i}-1}}({C_l},y_{{\bar\rho _i}-1}^l,\llbracket {{\mathbf{v}}_{{\bar\rho _i}-1}^l} 
% 				\rrbracket) \cap {\mathcal{B}_l^ - }({\mathbf{x}}_{{\bar\rho _i}-1})} \right)}  \hfill \\
% 		= \bigcap\limits_{l\in\mathcal{M}_{0}^i} \left({\mathcal{O}_{{\bar\rho _i}-1,{\bar\rho _i}-1}^l \cap \left( {A\left( {\mathcal{Q}}_{{\bar\rho _i}-2,0}^l \right) + B\llbracket {{{\mathbf{w}}_{{\bar\rho _i} - 2}}} 
% 				\rrbracket} \right)}\right) \hfill \\
%         \mathop  \subseteq \limits^{(b)}  \bigcap\limits_{l\in V_t^{\mathrm{s}}} \mathcal{O}_{0,0}^l \bigcap {\mathcal{Q}}_{{\bar\rho _i}-1,1}^l \hfill \\
%         =\mathcal{C}_{{\bar\rho_i }-1}^{i} \cap {\mathcal{Q}_{{\bar\rho_i }-1,1}^i}, \hfill
% 	\end{gathered}
% \end{equation}
% where $(b)$ holds since \eqref{eq_appproof_1} holds.
\begin{figure*}[t]%[bp] 
\begin{tcolorbox}[colback=gray!10, colframe=white, boxrule=0pt, arc=0pt]
\begin{align}\label{eq_appproof_1}
   % \begin{gathered}
        \bigcap\limits_{l\in\mathcal{M}_{0}^i} \left({\mathcal{O}_{{k'+1},{k'+1}}^l \bigcap \left( {A{{\mathcal{B}_l}({\mathbf{x}}_{k'})}  + B\llbracket {{{\mathbf{w}}_{k'}}} 
				\rrbracket} \right)}\right) &
		\mathop \subseteq \limits^{\eqref{eq_lm_minkintersect}} \bigcap\limits_{l\in\mathcal{M}_{0}^i}  \left({\mathcal{O}_{{k'+1},{k'+1}}^l \bigcap {\left(\bigcap\limits_{r = 0}^{k'} {\bigcap\limits_{j \in \mathcal{M}_{k' - r}^l} {\mathcal{O}_{k'+1,r}^j} }  \bigcap \bigcap\limits_{j \in \mathcal{M}_{k'}^l} {\mathcal{E}_{k'+1}^j}\right)} }\right)\hfill\notag\\
       & \mathop \subseteq \limits^{(a)} \bigcap\limits_{l\in\mathcal{M}_{0}^i}  \mathcal{O}_{{k'+1},{k'+1}}^l \bigcap\left(\bigcap\limits_{r = 0}^{k'} {\bigcap\limits_{l \in \mathcal{M}_{k'+1 - r}^i} {\mathcal{O}_{k'+1,r}^l}}\right)  \bigcap \left(\bigcap\limits_{l \in \mathcal{M}_{k'+1}^i} {\mathcal{E}_{k'+1}^l}\right)\notag\\
         &=\left(\bigcap\limits_{r = 0}^{k'+1} {\bigcap\limits_{l \in \mathcal{M}_{k'+1 - r}^i} {\mathcal{O}_{k'+1,r}^l}}\right)  \bigcap \left(\bigcap\limits_{l \in \mathcal{M}_{k'+1}^i} {\mathcal{E}_{k'+1}^l} \right)
    %\end{gathered}
\end{align}\end{tcolorbox}
\end{figure*}
% %

% \emph{Inductive step:}
% Assume \eqref{eq_prop_COITbound} for $k=k'$, when $k=k'+1$,we have,
% %
% \begin{equation}\label{eq_ap2_outb_2}
% 	\begin{gathered}
% 		\mathcal{B}_i({\mathbf{x}}_{k'+1}) = \bigcap\limits_{l\in\mathcal{M}_{0}^i} {\left( {{\mathcal{X}_{{k'+1}}}({C_l},y_{{k'+1}}^l,\llbracket {{\mathbf{v}}_{{k'+1}}^l} 
% 				\rrbracket) \cap {\mathcal{B}_l^ - }({\mathbf{x}}_{{k'+1}})} \right)}  \hfill \\
% 		= \bigcap\limits_{l\in\mathcal{M}_{0}^i} \left({\mathcal{O}_{{k'+1},{k'+1}}^l \cap \left( {A\left(\mathcal{C}_{k'}^{l}\cap {\mathcal{Q}}_{{k'},k'-{\bar\rho _i}+2}^l \right) + B\llbracket {{{\mathbf{w}}_{k'}}} 
% 				\rrbracket} \right)}\right) \hfill \\
%         \mathop  \subseteq \limits^{(b)}  \left(A\mathcal{C}_{k'}^{i} + B\llbracket {{{\mathbf{w}}_{k'}}} 
% 				\rrbracket\right)\bigcap \bigcap\limits_{l\in V_t^{\mathrm{s}}} \mathcal{O}_{{k'+1},{k'}-{\bar\rho _i}+2}^l\bigcap{\mathcal{Q}}_{{k'+1},{k'}-{\bar\rho _i}+3}^l\hfill \\
%         =\mathcal{C}_{{k'+1}}^{i} \cap {\mathcal{Q}_{{k'+1},{k'}-{\bar\rho _i}+3}^i}, \hfill
% 	\end{gathered}
% \end{equation}
% %
%
Thus, \propref{prop_COITbound} is proved by induction.
% From the mathematical induction, we know \eqref{eq_prop_COITbound} holds when $k\in\mathbb{N}_0$.
\hfill$\blacksquare$
% \section{Proof of \propref{prop_boundedCOIT}}\label{apx_prop_boundedCOIT}

% \hfill$\blacksquare$
%
\section{Proof of \propref{prop_decompCOIT}}\label{apx_decompo_COITb}
To start with, we provide the following lemma to support our proof. 
\begin{lemma}[An Outer Bound w.r.t. Block Matrix \cite{cong2022stability}]\label{lm_blockmatrixouterbound}
	For \[H =  {\begin{bmatrix}
			{{H_{11}}}& \cdots &{{H_{1q}}} \\ 
			\vdots & \ddots & \vdots  \\ 
			{{H_{p1}}}& \cdots &{{H_{p1}}} 
	\end{bmatrix}},\quad \mathcal{S} = {\mathcal{S}_1} \times  \cdots  \times {\mathcal{S}_q},\] 
	where ${H_{ij}} \in {\mathbb{R}^{{m_i} \times {n_j}}}$ and ${\mathcal{S}_j} \subset {\mathbb{R}^{{n_j}}}$, the following holds
	\begin{equation}\label{eq_lm_blockbound}
		H\mathcal{S} \subseteq \prod\limits_{i = 1}^p {\left( {\sum\limits_{j = 1}^q {{H_{ij}}{\mathcal{S}_j}} } \right)}.
	\end{equation}
\end{lemma}
\begin{Proof}
    For all $t\in H\mathcal{S}$, there exists $$\underbrace{\textnormal{col}(s_1,\dots,s_q)}_{s}\in\prod_{i = 1}^q\mathcal{S}_i$$ such that $t=Hs$, i.e., $$\underbrace{\textnormal{col}\left(\sum_{j = 1}^q {{H_{1j}}s_j, \ldots ,} \sum_{j = 1}^q {{H_{pj}}s_j}\right)}_{t}\in\prod_{i = 1}^p {(\sum_{j = 1}^q {{H_{ij}}{S_j}} )}.$$
    Thus, \eqref{eq_lm_blockbound} is derived.
\end{Proof}
From \eqref{eq_coit}, we have $\mathcal{B}_i({\mathbf{x}}_k)\subseteq\mathcal{C}_k^{(t)}$ for $k >\tilde{\rho}$, which gives
\begin{equation}\label{eq_proofapxb_ckt}
     P_{o}^{(t)}\mathcal{B}_i({\mathbf{x}}_k)\subseteq P_o^{(t)}\mathcal{C}_k^{(t)}.
\end{equation}
Thus, to derive~\eqref{eq_propdecopCOIT_outbo}, we only need to prove $P_{\bar o}^{(t)}\mathcal{B}_i({\mathbf{x}}_k) \subseteq \mathcal{\tilde S}_k^{i}$ which is provided as follows.\\
For $k=0$, according to \algref{alg:thmframe}, one has
\begin{equation}\label{eq_proof3_basecase}
    \begin{split}
	P_{\bar o}^{(t)}\mathcal{B}_i({\mathbf{x}}_0) 
    &\mathop  = \limits^{\eqref{eq_distrframe1}} P_{\bar o}^{(t)}\bigcap\limits_{j\in{\mathcal{M}_{0}^i}} {{{\mathcal{B}_j^ + }({\mathbf{x}}_0)}}  \\ 
    &\mathop  = \limits^{\eqref{eq_distrframe3}} P_{\bar o}^{(t)} \bigcap\limits_{j\in{\mathcal{M}_{0}^i}} {{\mathcal{B}_j^ - }({\mathbf{x}}_0)}\\
    &\subseteq \underbrace{P_{\bar o}^{(t)}{\mathcal{B}_i^ - }({\mathbf{x}}_0)}_{\mathcal{\tilde S}_0^{i}}.
    \end{split}
\end{equation}
For $k=1$, we have
\begin{equation}\label{eq_proofapxb1}
	\begin{split}
		P_{\bar o}^{(t)}\mathcal{B}_i({\mathbf{x}}_1) &= P_{\bar o}^{(t)}\bigcap\limits_{j\in{\mathcal{M}_{0}^i}} {{{\mathcal{B}_j^ + }({\mathbf{x}}_1)}} 
        \\&\mathop= \limits^{\eqref{eq_distrframe2}} P_{\bar o}^{(t)}\bigcap\limits_{j\in{\mathcal{M}_{0}^i}}{\mathcal{B}_j^ - }({\mathbf{x}}_1)
        \\& \subseteq P_{\bar o}^{(t)}{\mathcal{B}_i^ - }({\mathbf{x}}_1) 
       \\& =P_{\bar o}^{(t)}({A\mathcal{B}_i({\mathbf{x}}_0) + B\llbracket {{{\mathbf{w}}_0}}\rrbracket})
        \\&= P_{\bar o}^{(t)}A ({P^{(t)}})^{-1}P^{(t)}\mathcal{B}_i({\mathbf{x}}_0) + P_{\bar o}^{(t)}B\llbracket {{{\mathbf{w}}_0}}\rrbracket.   
	\end{split}
\end{equation}
Then, basd on \eqref{eq_proof3_basecase}, we can rewrite \eqref{eq_proofapxb1} as
\begin{equation}\label{eq_proofapxb2}
	\begin{split}
		P_{\bar o}^{(t)}\mathcal{B}_i({\mathbf{x}}_1) 
        & \subseteq \begin{bmatrix}
         \tilde A_{21}^{(t)}&\tilde A_{\bar o}^{(t)}
         \end{bmatrix}\big({(P_{o}^{(t)} \mathcal{B}_i(\mathbf{x}_0)) \times \mathcal{\tilde S}_0^{i}}\big)+ P_{\bar o}^{(t)}B\llbracket {{{\mathbf{w}}_0}}\rrbracket  \\
        &\!\!\!\!\!\!\! \mathop\subseteq \limits^{\rm\lemref{lm_blockmatrixouterbound}} {\tilde A_{\bar o}^{(t)}\mathcal{\tilde S}_0^{i} + \tilde A_{21}^{(t)}P_{o}^{(t)} \mathcal{B}_i(\mathbf{x}_0) + \tilde B_{\bar o}^{(t)}\llbracket {{{\mathbf{w}}_0}}\rrbracket}\\
       & = \mathcal{\tilde S}_1^{i}.   
	\end{split}
\end{equation}
%
% where $(a)$ holds since
% \begin{equation}
%      {\mathcal{B}_i^ - }({\tilde{\mathbf{x}}}_1^{{{\bar o}}})=\begin{bmatrix}
%          \tilde A_{\bar o}^{(t)}&\tilde A_{21}^{(t)}
%      \end{bmatrix}
%      \begin{bmatrix}
%          \mathcal{B}_i^-({\tilde{\mathbf{x}}}_0^{{{\bar o}}})\\
%          \mathcal{\tilde C}_0^{i}
%      \end{bmatrix}+ \tilde B_{\bar o}^{(t)}\llbracket {{{\mathbf{w}}_0}} 
% 			\rrbracket,
% \end{equation}
% %
% according to \eqref{eq_distrframedec1}, and the property demonstrated in \eqref{eq_lm_blockbound}.
% %
Proceeding forward, when $k > 0$, we have 
\begin{equation}
    P_{\bar o}^{(t)}{\mathcal{B}_i}({\mathbf{x}}_k) \subseteq \mathcal{\tilde S}_k^{i},
\end{equation}
which, upon combining with \eqref{eq_proofapxb_ckt}, gives~\eqref{eq_propdecopCOIT_outbo}.
% \begin{equation}
%         {P^{(t)}}{\mathcal{B}_i}({\mathbf{x}}_k)=\mathcal{\tilde C}_k^{(t)} \times \mathcal{\tilde S}_k^{i}.
% \end{equation}
% when $k \geqslant {\bar\rho}-1$, we have $\mathcal{B}_i({\tilde{\mathbf{x}}}_k^{o}) \subseteq \mathcal{\tilde C}_k^{(t)}$, which gives  
% %
% \begin{multline}
%     \mathcal{B}_i({\tilde{\mathbf{x}}}_k^{\bar o})\subseteq \mathcal{\tilde S}_k^{{i}} =  (\tilde A_{\bar o}^{(t)})^k\mathcal{B}_i^-({{\tilde{\mathbf{x}}}_{0}^{\bar o}} 
% 			) \\+ \sum\limits_{j = {\bar\rho-1}}^{k - 1} {(\tilde A_{\bar o}^{(t)})^{k - 1 - j}({\tilde A_{21}^{(t)}}\mathcal{\tilde C}_j^{(t)} + {\tilde B_{\bar o}^{(t)}}\llbracket {{{\mathbf{w}}_j}} 
% 				\rrbracket)} \\ + \sum\limits_{j = 0}^{{\bar\rho} - 2} {(\tilde A_{\bar o}^{(t)})^{k - 1 - j}(\tilde A_{21}^{(t)} \mathcal{\tilde C}_j^{i} + \tilde B_{\bar o}^{(t)}\llbracket {{{\mathbf{w}}_j}} 
% 				\rrbracket)}.
% \end{multline}
%
% Proceeding forward, we know $\mathcal{B}_i({\tilde{\mathbf{x}}}_k^{{{\bar o}}}) \subseteq \mathcal{\tilde S}_k^{i}$ holds for $k \in {\mathbb{N}_0}$.
% %
% Therefore, it is concluded that $\mathcal{B}_i( {{\tilde{\mathbf{x}}}_{k}} 
% ) \subseteq  \mathcal{\tilde C}_k^{(t)} \times \mathcal{\tilde S}_k^{i}$ holds for $k > {\bar\rho}$. 
% %
\hfill$\blacksquare$

\section{Proof of \thmref{thm_boundednesssuffcient}}\label{apx_thmoundednesssuffcient}

To start with, we provide two lemmas and a proposition.

\begin{lemma}[\!\cite{9031321}]\label{lem_normmatrix}
    Given $A\in \mathbb{R}^{n\times n}$, for each $\gamma>{\rm Rho}(A)$ (${\rm Rho}(A)$ stands for the spectral radius of $A$), there exists a constant $c_\gamma\geq 1$, such that $\forall k\in \mathbb{N}_0$, $\left\| {{F^k}} \right\| \leqslant {c_\gamma }{\gamma ^k}$.
\end{lemma}

\begin{lemma}\label{lm_unniformboundedinters}
    Let $\mathcal{J} \subseteq \mathbb{N}_0$ be finite, and suppose that $$\sup_{k \geqslant \underline{k}} D(\mathcal{S}_k^j)<\infty, \quad \forall j \in \mathcal{J},$$ where $\underline{k} \geq 0$.
    Define $F = \textnormal{col}({F_{\min\mathcal{J}}}, \ldots ,{F_{\max\mathcal{J}}})$ and $F_j \in {\mathbb{R}^{q_j \times n_o}}$, if ${\mathrm{rank}}(F) = n_o$,
	% \begin{equation}\label{eq_lmecodnition}
	% 	{\mathrm{rank}}(F) = n_o,
	% \end{equation}
	%
	the following holds:
	\begin{equation}\label{eq_lm_uniformlyboundedinters}
		\mathop {\sup }\limits_{k \geqslant \underline{k}} D\bigg(\bigcap\limits_{j \in \mathcal{J}}  {[\ker (F_j) + \mathcal{S}_k^j]}\bigg)<\infty.
	\end{equation}	
\end{lemma}

\begin{Proof}
    It directly follows from Lemma~3 in~\cite{cong2022stability}.
\end{Proof}

\begin{proposition}\label{prop_boundedCOIT}
    $\forall G_t^{\mathrm{s}} \subseteq G$, the following holds
    \begin{equation}\label{eqn:Bounded Projection}
        \sup_{k \geq \tilde{\rho} + \nu - 2} D\Big(P_o^{(t)} \mathcal{C}_k^{(t)}\Big) < \infty,
    \end{equation}
    where $\nu$ is the observability index\cite{Chen1984} of $({{{\tilde A}_{o}^{(t)}}}, {\tilde C_{o}^{(t)}})$.
		% $\forall G_t^{\mathrm{s}} \subseteq G$, if the system pair $(A,C^{(t)})$ is observable, the COIT defined by \eqref{eq_coit} satisfies \[\mathop {\overline {\lim } }\limits_{k \to \infty } D(\mathcal{C}_k^{(t)})<\infty.\] 
  %       % where $\mu$ denotes the observability index\footnote{More specifically, for system pair $(A,C)$, the observability index $\mu$ is the smallest $v$ such that the matrix $O_v = \begin{bmatrix} (C)^\top& \ldots &(CA^v)^\top \end{bmatrix}^\top$ has full column rank, i.e., $\mu = \arg\min_{v} \{O_v\colon \mathrm{rank}(O_v) = n_o\}$. A detailed explanation can be found in \cite{Chen1984}.} of $(A,C^{(t)})$.
\end{proposition}
\begin{Proof}
$\forall G_t^{\mathrm{s}} \subseteq G$, with~\eqref{eq_ker_dsf},~\eqref{eq_observationinformationtdistri1}, and~\eqref{eq_coit}, we have
\begin{equation}\label{eqn:Outer Bound of Projected COIT}
    \begin{split}
        P_o^{(t)} \mathcal{C}_k^{(t)} &\subseteq \bigcap\limits_{r = 0}^{k-\tilde{\rho}+1}  {\bigcap\limits_{l\in V_t^{\mathrm{s}}} {\Big[ P_o^{(t)} {A^{k- r}}\ker ({C_l}) +   \mathcal{S}_k^{(r,l)}} \Big]}\\
        &\stackrel{(b)}{\subseteq} \bigcap\limits_{r = \underline{k}}^{k-\tilde{\rho}+1}  {\bigcap\limits_{l\in V_t^{\mathrm{s}}} {\Big[ \ker ({\tilde{C}_{o,l}^{(t)}} \big(\tilde A_{o}^{(t)}\big)^{r-k}) + \mathcal{S}_k^{(r,l)}} \Big]},
    \end{split}
\end{equation}
where $\underline{k} =\underbrace{ k-\tilde{\rho}-\nu+2}_{\geq 0} $, ${\tilde{C}_{o,l}^{(t)}} = C_l (P^{(t)})^{-1} \begin{bmatrix}I_{n_o}\\0\end{bmatrix}$, and
\begin{equation*}%\label{eq_COIT_Sboundedpart}
	\mathcal{S}_k^{(r,l)} = P_o^{(t)} \bigg[{A^{k - r}}C_l^\dag \Big(\{ y_{r}^l\}  + \llbracket { - {\mathbf{v}}_{r}^l} 
	\rrbracket\Big) + \sum\limits_{\tau = r}^{k-1} {A^{k -1 -\tau}}B\llbracket {{\mathbf{w}_\tau}\rrbracket}\bigg].
\end{equation*}
In~\eqref{eqn:Outer Bound of Projected COIT},~$(b)$ follows from
\begin{equation*}
    P_o^{(t)} {A^{k- r}} = \big(\tilde A_{o}^{(t)}\big)^{k-r} P_o^{(t)},\quad P_o^{(t)} \ker ({C_l}) = \ker ({\tilde{C}_{o,l}^{(t)}}),
\end{equation*}
and $$\big(\tilde A_{o}^{(t)}\big)^{k-r} \ker ({\tilde{C}_{o,l}^{(t)}}) = \ker \Big({\tilde{C}_{o,l}^{(t)}} \big(\tilde A_{o}^{(t)}\big)^{r-k}\Big).$$
%
% \begin{multline}\label{eq_COIT_ker}
% 	\mathcal{C}_k^{(t)} = \bigcap\limits_{r = 0}^{k-\bar\rho+1}  {\bigcap\limits_{l\in G_t^{\mathrm{s}}} {({A^{k- r}}\ker ({C_l}) + \mathcal{S}_{r,k}^l} )}\\
%     = \bigcap\limits_{r = 0}^{k-\bar\rho+1}  {\bigcap\limits_{l\in G_t^{\mathrm{s}}} {(\ker ({C_l}{A^{r-k}}) + \mathcal{S}_{r,k}^l})},
% \end{multline}
%
%
%
% By ${A^{k  - r}}\ker (C_l ) = {A^{k- r}}R({{\bar C}_l^ \top}) = R({A^{k - r}}{{\bar C}_l^ \top}) = :R({{\bar F}_{l,r}})$, \eqref{eq_COIT_ker} can be rewritten as 
% \begin{equation}\label{eq_COIT_enhaneceboundedpart2}
% 	\mathcal{C}_k^{(t)}=\bigcap\limits_{r = 0}^{k-\bar\rho+1}  {\bigcap\limits_{l\in G_t^{\mathrm{s}}}{(R({{\bar F}_{l,r}}) + \mathcal{S}_{r,k}^l} )}.
% \end{equation}
% %
With~\eqref{eqn:Uniformly Bounded Noises} and $\underline{k} \leqslant r \leqslant k-\tilde{\rho}+1$, we get $$\sup_{k \geqslant \underline{k}} D(\mathcal{S}_k^{(r,l)})<\infty.$$
Let $j = r|V| + l$ and $F_j = F_{(r,l)} = {\tilde{C}_{o,l}^{(t)}} [\tilde A_{o}^{(t)}]^{r-k}$.
We group those $F_j = F_{(r,l)}$ with the same $r$ together, and get $$F_{(r)} = \textnormal{col}(F_{(r, \min V_t^{\mathrm{s}})},\dots,F_{(r, \max V_t^{\mathrm{s}})}) = \tilde{C}_o^{(t)} [\tilde A_{o}^{(t)}]^{r-k}.$$
Thus, $F = \textnormal{col}({F_{\min\mathcal{J}}}, \ldots ,{F_{\max\mathcal{J}}})$, with $$\min\mathcal{J} = \underline{k}|V| + \min V_t^{\mathrm{s}}$$ and $$\max\mathcal{J} = (k-\tilde{\rho}+1)|V| + \max V_t^{\mathrm{s}},$$ has the following form:
\begin{equation*}
    F =
    \underbrace{\begin{bmatrix}
        {\tilde{C}_{o,l}^{(t)}}\\
        \vdots\\
        {\tilde{C}_{o,l}^{(t)}} \big(\tilde A_{o}^{(t)}\big)^{\nu-1}
    \end{bmatrix}}_{\tilde{O}_{\nu}} \big(\tilde A_{o}^{(t)}\big)^{\underline{k}-k} =: \tilde{O}_{\nu} \big(\tilde A_{o}^{(t)}\big)^{\underline{k}-k}.
\end{equation*}
Since $\nu$ is the observability index of $({{{\tilde A}_{o}^{(t)}}}, {\tilde C_{o}^{(t)}})$, we have $\mathrm{rank}(\tilde{O}_{\nu}) = n_o$, and hence $\mathrm{rank}(F) = n_o$.
%
% %
% Let $F_{l,r}^ \top : = {C_l}{A^{r - k}}$,  when $k\to\infty$, we have
% \[\begin{split}
% 	&{\rm rank}({F^ \top })
% 	\\&={\rm rank}({\begin{bmatrix}
% 			{F_{i_{t,1},0}^ \top } \\ 
% 			\vdots  \\ 
% 			{F_{i_{t,1},r}^ \top } \\ 
% 			\vdots  \\ 
% 			{F_{i_{t,m},k- {\bar\rho}+1}^ \top } 
% 	\end{bmatrix}}) = {\rm rank}( { {\begin{bmatrix} 
% 				C^{(t)}  \\ 
% 				{C^{(t)}A} \\ 
% 				\vdots  \\ 
% 				{C^{(t)}{A^{k-{\bar\rho}+1} }} 
% 		\end{bmatrix}}{A^{ - k}}})\mathop  = \limits^{(b)} n,
% \end{split}\]
% where $(b)$ holds since pair $(A,C^{(t)})$ is observable.
%
Therefore, by~\eqref{eqn:Outer Bound of Projected COIT} and \lemref{lm_unniformboundedinters},~\eqref{eqn:Bounded Projection} is satisfied.
%for a source component $G_t^{\mathrm{s}}$ of $G$, the COIT satisfies $\mathop {\overline {\lim } }\nolimits_{k \to \infty } D(\mathcal{C}_k^{(t)})<\infty$.
\end{Proof}
Now, we prove \thmref{thm_boundednesssuffcient}, which includes two steps.
In the first step, we show that~\eqref{eq_boundedness_definition} holds [i.e., $\mathop {\overline {\lim } }\nolimits_{k \to \infty } D(\mathcal{B}_i({\mathbf{x}}_{k}))<\infty$] for the sensors in the source components;
in the second step,~\eqref{eq_boundedness_definition} is proven for the other sensors.
% %
% First, the boundedness of nodes that are included in any source components is analyzed; second, the boundedness of nodes that are not included in any source components is analyzed.
% %

\emph{Step~1:}
From \propref{prop_decompCOIT}, $\forall i \in \bigcup_{t=1}^T V_t^{\mathrm{s}}$,
%$P^{(t)}\mathcal{B}_i( {{\mathbf{x}}_{k}})$ is outerbounded by ${P_o^{(t)} \mathcal{C}_k^{(t)} \times \mathcal{\tilde S}_k^{i}}$ for $k >{\bar\rho}$.
%
we have
\begin{align}\label{eqinpf:thm_boundednesssuffcient - Step 1 - Total Bound}
    \varlimsup_{k\to\infty} D(\mathcal{B}_i({\mathbf{x}}_k)) \leqslant &\ \big\|(P^{(t)})^{-1}\big\| \varlimsup_{k\to\infty} D(P_o^{(t)} \mathcal{C}_k^{(t)}) \notag\\
    &+ \big\|(P^{(t)})^{-1}\big\| \varlimsup_{k\to\infty} D(\mathcal{\tilde S}_k^{i}).
\end{align}
%
% \begin{equation*}
%     \varlimsup_{k\to\infty} D(P^{(t)}\mathcal{B}_i({\mathbf{x}}_k)) \leqslant \varlimsup_{k\to\infty} D(P_o^{(t)} \mathcal{C}_k^{(t)}) + \varlimsup_{k\to\infty} D(\mathcal{\tilde S}_k^{i}).
% \end{equation*}
%
From \propref{prop_boundedCOIT}, we know that $$\varlimsup_{k\to\infty} D(P_o^{(t)} \mathcal{C}_k^{(t)}) < \infty.$$
Thus, to prove~\eqref{eq_boundedness_definition}, we only focus on $\varlimsup_{k\to\infty} D(\mathcal{\tilde S}_k^{i})$ in~\eqref{eqinpf:thm_boundednesssuffcient - Step 1 - Total Bound}.
With~\eqref{eq_prop2_sij}, we divide $\mathcal{\tilde S}_k^{i}$ into three parts
\begin{equation}\label{eqinpf:thm_boundednesssuffcient - Step 1 - Three Parts of S}
    \mathcal{\tilde S}_k^{i} = \mathcal{\tilde S}_{k,1}^{i} + \mathcal{\tilde S}_{k,2}^{i} + \mathcal{\tilde S}_{k,3}^{i},
\end{equation}
where
\begin{align}
    \mathcal{\tilde S}_{k,1}^{i} &= (\tilde A_{\bar o}^{(t)})^k P_{\bar{o}}^{(t)} \mathcal{B}_i^-(\mathbf{x}_0),\label{eqinpf:thm_boundednesssuffcient - Step 1 - Three Parts of S - Part 1} \\ 
    \mathcal{\tilde S}_{k,2}^{i} &=\!\!\!\!\sum\limits_{j = 0}^{\tilde{\rho} + \nu - 3} {(\tilde A_{\bar o}^{(t)})^{k - 1 - j}({\tilde A_{21}^{(t)}} P_{o}^{(t)} \mathcal{B}_i(\mathbf{x}_j) + {\tilde B_{\bar o}^{(t)}}\llbracket {{{\mathbf{w}}_j}}\rrbracket)}, \label{eqinpf:thm_boundednesssuffcient - Step 1 - Three Parts of S - Part 2}\\
    \mathcal{\tilde S}_{k,3}^{i} &\mathop \subseteq \limits^{\eqref{eq_proofapxb_ckt}}\!\!\!\!\!\sum\limits_{j = \tilde{\rho} + \nu - 2}^{k - 1} {(\tilde A_{\bar o}^{(t)})^{k - 1 - j}({\tilde A_{21}^{(t)}} P_{o}^{(t)} \mathcal{C}_j^{(t)} + {\tilde B_{\bar o}^{(t)}}\llbracket {{{\mathbf{w}}_j}}\rrbracket)}.\label{eqinpf:thm_boundednesssuffcient - Step 1 - Three Parts of S - Part 3}
\end{align}
Since ${{\tilde A}_{\bar o}^{(t)}}$ is marginally stable [condition~(i) in \thmref{thm_boundednesssuffcient}], $$\varlimsup_{k\to\infty} \|(\tilde A_{\bar o}^{{(t)}})^k\| < \infty,$$
which together with $D(\mathcal{B}_i^-({{{\mathbf{x}}}_{0}}))<\infty$,~\eqref{eqn:Uniformly Bounded Noises}, and \algref{alg:thmframe} gives
\begin{align}\label{eqinpf:thm_boundednesssuffcient - Step 1 - Asymptotically Bounded S1 and S2}
    \varlimsup_{k\to\infty} D(\mathcal{\tilde S}_{k,1}^{i}) < \infty, & &\varlimsup_{k\to\infty} D(\mathcal{\tilde S}_{k,2}^{i}) < \infty.
\end{align}
Now, we focus on $\mathcal{\tilde S}_{k,3}^{i}$.
For marginally stable ${{\tilde A}_{\bar o}^{(t)}}$, its eigenvalues satisfy $|\lambda| \leq 1$, and those with $|\lambda| = 1$ (namely, $\lambda_{1}^{=1},\dots,\lambda_{p}^{=1}$) are semisimple.
Let $q_1,\ldots, q_p$ be the left eigenvectors associated with $\lambda_{1}^{=1},\dots,\lambda_{p}^{=1}$.
Under condition~(ii) of \thmref{thm_boundednesssuffcient}, $\forall j \in \{1,\ldots,p\}$, the following holds
\begin{equation}\label{eq_thmfabprof_1}
	{q_j}{\begin{bmatrix}
			{{{{{\tilde A}_{\bar o}^{(t)}}}} - \lambda_j^{=1}I_{n_{\bar{o}}}}&{{{{\tilde B}_{\bar o}^{(t)}}}}&{{{{\tilde A}_{21}^{(t)}}}} 
	\end{bmatrix}} = 0.
\end{equation}
Thus, with $Q_{=1} := \textnormal{col}(q_1,\ldots,q_p)$, we have
\begin{align}\label{eqinpf:thm_boundednesssuffcient - Step 1 - transformation - part 1}
    Q_{=1} {{{{\tilde A}_{\bar o}^{(t)}}}} &= \Lambda_{=1} Q_{=1},\notag\\
    \quad Q_{=1} {{\tilde B}_{\bar o}^{(t)}}& = 0, \\
    \quad Q_{=1} {{\tilde A}_{21}^{(t)}} &= 0,\notag
\end{align}
where $\Lambda_{=1} = \mathrm{diag}(\lambda_{1}^{=1},\dots,\lambda_{p}^{=1})$.
This means that we can find $Q_{=1}$ such that
\begin{align}\label{eqinpf:thm_boundednesssuffcient - Step 1 - transformation - part 2}
    Q = \begin{bmatrix}Q_{<1}\\Q_{=1}\end{bmatrix}, & & {{\tilde A}_{\bar o}^{(t)}} = Q^{-1}
    \begin{bmatrix}
        J_{<1} & \\
        & \Lambda_{=1}
    \end{bmatrix} Q,
\end{align}
where $\mathrm{Rho} (J_{<1}) < 1$.
%
% \begin{equation*}
%     Q = \begin{bmatrix}Q_{=1}\\Q^{<1}\end{bmatrix}, \quad {{\tilde A}_{\bar o}^{(t)}} = Q^{-1}
%     \begin{bmatrix}
%         J^{<1} & 0_{(n_{\bar{o}} - p) \times p}\\
%         0_{p \times (n_{\bar{o}} - p)} & \Lambda_{=1}
%     \end{bmatrix} Q.
% \end{equation*}

With~\eqref{eqinpf:thm_boundednesssuffcient - Step 1 - transformation - part 1} and~\eqref{eqinpf:thm_boundednesssuffcient - Step 1 - transformation - part 2}, we can rewrite the RHS of~\eqref{eqinpf:thm_boundednesssuffcient - Step 1 - Three Parts of S - Part 3} as
\begin{equation*}
    Q^{-1} \sum\limits_{j = \tilde{\rho} + \nu - 2}^{k - 1}
    \begin{bmatrix}
        J_{<1}^{k-1-j} Q_{<1} ({\tilde A_{21}^{(t)}} P_{o}^{(t)} \mathcal{C}_j^{(t)} + {\tilde B_{\bar o}^{(t)}}\llbracket {{{\mathbf{w}}_j}}\rrbracket)\\
        0
    \end{bmatrix},
\end{equation*}
which means
\begin{align}\label{eqinpf:thm_boundednesssuffcient - Step 1 - limsup DS3}
    \varlimsup_{k\to\infty} D(\mathcal{\tilde S}_{k,3}^{i}) \leqslant & \varlimsup_{k\to\infty} \|Q^{-1}\| \sum\limits_{j = \tilde{\rho} + \nu - 2}^{k - 1} \| J_{<1}^{k-1-j}\| \|Q_{<1} \| \notag \\
   & \times \Big[\|{\tilde A_{21}^{(t)}}\| D(P_{o}^{(t)}\mathcal{C}_j^{(t)}) + \|{\tilde B_{\bar o}^{(t)}}\| D(\llbracket {{{\mathbf{w}}_j}}\rrbracket)\Big]  \notag \\
   \stackrel{(c)}{<} &\infty.
\end{align}
where $(c)$ follows from $\varlimsup_{k\to\infty} D(P_o^{(t)} \mathcal{C}_k^{(t)}) < \infty$ (guaranteed by \propref{prop_boundedCOIT}), equation~\eqref{eqn:Uniformly Bounded Noises}, and \lemref{lem_normmatrix}.

Equation~\eqref{eqinpf:thm_boundednesssuffcient - Step 1 - Three Parts of S} together with~\eqref{eqinpf:thm_boundednesssuffcient - Step 1 - Asymptotically Bounded S1 and S2} and~\eqref{eqinpf:thm_boundednesssuffcient - Step 1 - limsup DS3} gives $\varlimsup_{k\to\infty} D(\mathcal{\tilde S}_k^{i}) < \infty$ in~\eqref{eqinpf:thm_boundednesssuffcient - Step 1 - Total Bound}.
Therefore,
\begin{align}\label{eqinpf:thm_boundednesssuffcient - Step 1 - Total Bound - End}
    \varlimsup_{k\to\infty} D(\mathcal{B}_i({\mathbf{x}}_k)) < \infty, & & \forall i \in \bigcup_{t=1}^T V_t^{\mathrm{s}}.
\end{align}

\emph{Step~2:}
If sensor $i$ is not included in any source
components, it can still receive a bounded local posterior belief from the sensor in source components. 
%
% Let sensor $i$ in $\rho_{sc}$-hop predecessors of sensor $i'$ in a source component. 
% %
According to \eqref{eq_distrframe1}-\eqref{eq_distrframe3}, we have 
\begin{equation}\label{eq_proof4_step2_1}
        \mathcal{B}_i({\mathbf{x}}_{k}) \subseteq  {A{\mathcal{B}_j}({\mathbf{x}}_{k-1})}+B\llbracket {{{\mathbf{w}}_{k-1}}}\rrbracket,
\end{equation}
where $j\in \mathcal{M}_0^i$.
Since there exists a sensor $i'$ that belongs to a source component $G_{t}^{\rm s}$ has a direct path to $i$, letting this sensor be the $\rho_{sc}$-hop predecessor of $i$, we have
\begin{equation}\label{key}
        \mathcal{B}_i({\mathbf{x}}_{k}) \subseteq {{A^{\rho_{sc}}{\mathcal{B}_{i'}}({\mathbf{x}}_{k-\rho_{sc}})}+\sum\limits_{r = {1}}^{\rho_{sc}} {A^{r-1}}B\llbracket {{{\mathbf{w}}_{k-\rho_{sc}}}}\rrbracket},
\end{equation}
by applying \eqref{eq_proof4_step2_1} recursively.
Since $\varlimsup_{k\to\infty} D(\mathcal{B}_{i'}({\mathbf{x}}_k)) < \infty$ according to \eqref{key}, we have
\begin{align}\label{eq_boundedness_notin}
        \varlimsup_{k\to\infty} D(\mathcal{B}_i({\mathbf{x}}_{k})) \leq &  \varlimsup_{k\to\infty} \Big[\|A^{\rho_{sc}}\|D({\mathcal{B}_{i'}}({\mathbf{x}}_{k-\rho_{sc}}))\notag\\
        &+\sum\limits_{r = {1}}^{\rho_{sc}} \|{A^{r-1}}B\|D(\llbracket {{{\mathbf{w}}_{k-\rho_{sc}}}}\rrbracket)\Big]\notag\\
        <&\ \infty.
\end{align}

Based on \eqref{eqinpf:thm_boundednesssuffcient - Step 1 - Total Bound - End} and \eqref{eq_boundedness_notin}, \algref{alg:thmframe} can achieve a bounded estimate for all the sensors in $G$.
\hfill$\blacksquare$	
\bibliographystyle{IEEEtran}	
\bibliography{ww}

\begin{thebibliography}{10}
\providecommand{\url}[1]{#1}
\csname url@samestyle\endcsname
\providecommand{\newblock}{\relax}
\providecommand{\bibinfo}[2]{#2}
\providecommand{\BIBentrySTDinterwordspacing}{\spaceskip=0pt\relax}
\providecommand{\BIBentryALTinterwordstretchfactor}{4}
\providecommand{\BIBentryALTinterwordspacing}{\spaceskip=\fontdimen2\font plus
\BIBentryALTinterwordstretchfactor\fontdimen3\font minus
  \fontdimen4\font\relax}
\providecommand{\BIBforeignlanguage}[2]{{%
\expandafter\ifx\csname l@#1\endcsname\relax
\typeout{** WARNING: IEEEtran.bst: No hyphenation pattern has been}%
\typeout{** loaded for the language `#1'. Using the pattern for}%
\typeout{** the default language instead.}%
\else
\language=\csname l@#1\endcsname
\fi
#2}}
\providecommand{\BIBdecl}{\relax}
\BIBdecl

\bibitem{8425647}
Y.~Wang, Z.~Wang, V.~Puig, and G.~Cembrano, ``Zonotopic set-membership state
  estimation for discrete-time descriptor lpv systems,'' \emph{IEEE Trans.
  Autom. Control}, vol.~64, no.~5, pp. 2092--2099, 2019.

\bibitem{cong2022stability}
Y.~Cong, X.~Wang, and X.~Zhou, ``Stability of linear set-membership filters
  with respect to initial conditions: An observation-information perspective,''
  \emph{Automatica}, vol. 172, p. 111993, 2025.

\bibitem{10750409}
Y.~Li, Y.~Cong, and J.~Dong, ``Existence and completeness of bounded
  disturbance observers: A set-membership viewpoint,'' \emph{IEEE Trans. Autom.
  Control}, pp. 1--8, 2024.

\bibitem{garcia2020guaranteed}
R.~A. Garc{\'\i}a, L.~Orihuela, P.~Mill{\'a}n, F.~R. Rubio, and M.~G. Ortega,
  ``Guaranteed estimation and distributed control of vehicle formations,''
  \emph{International Journal of Control}, vol.~93, no.~11, pp. 2729--2742,
  2020.

\bibitem{orihuela2017distributed}
L.~Orihuela, S.~Roshany-Yamchi, R.~A. Garc{\'\i}a, and P.~Mill{\'a}n,
  ``Distributed set-membership observers for interconnected multi-rate
  systems,'' \emph{Automatica}, vol.~85, pp. 221--226, 2017.

\bibitem{ierardi2021distributed}
C.~Ierardi, L.~Orihuela, and I.~Jurado, ``A distributed set-membership
  estimator for linear systems with reduced computational requirements,''
  \emph{Automatica}, vol. 132, p. 109802, 2021.

\bibitem{ZHU2025112345}
``Mean-shift-based robust distributed set-membership fusion filtering for
  sensor network systems with outliers,'' \emph{Automatica}, vol. 177, p.
  112345, 2025.

\bibitem{alanwar2023distributed}
A.~Alanwar, J.~J. Rath, H.~Said, K.~H. Johansson, and M.~Althoff, ``Distributed
  set-based observers using diffusion strategies,'' \emph{Journal of the
  Franklin Institute}, vol. 360, no.~10, pp. 6976--6993, 2023.

\bibitem{Liu2021}
L.~Liu, L.~Ma, J.~Guo, J.~Zhang, and Y.~Bo, ``Distributed set-membership
  filtering for time-varying systems: A coding–decoding-based approach,''
  \emph{Automatica}, vol. 129, p. 109684, 2021.

\bibitem{XIE2025112347}
``Set-membership estimator design with privacy-preserving for sensor networks:
  A state-decomposition-based approach,'' \emph{Automatica}, vol. 177, p.
  112347, 2025.

\bibitem{8793114}
D.~Ding, Z.~Wang, and Q.-L. Han, ``A set-membership approach to event-triggered
  filtering for general nonlinear systems over sensor networks,'' \emph{IEEE
  Trans. Autom. Control}, vol.~65, no.~4, pp. 1792--1799, 2020.

\bibitem{8605375}
S.~Liu, Z.~Wang, G.~Wei, and M.~Li, ``Distributed set-membership filtering for
  multirate systems under the round-robin scheduling over sensor networks,''
  \emph{IEEE Trans. Cybern.}, vol.~50, no.~5, pp. 1910--1920, 2020.

\bibitem{7463019}
S.~Park and N.~C. Martins, ``Design of distributed {LTI} observers for state
  omniscience,'' \emph{IEEE Trans. Autom. Control}, vol.~62, no.~2, pp.
  561--576, 2017.

\bibitem{wang2024distributed}
S.~Wang and M.~Guay, ``Distributed state estimation for jointly observable
  linear systems over time-varying networks,'' \emph{Automatica}, vol. 163, p.
  111564, 2024.

\bibitem{BATTISTELLI2014707}
G.~Battistelli and L.~Chisci, ``Kullback–leibler average, consensus on
  probability densities, and distributed state estimation with guaranteed
  stability,'' \emph{Automatica}, vol.~50, no.~3, pp. 707--718, 2014.

\bibitem{8845692}
W.~Li, Z.~Wang, D.~W.~C. Ho, and G.~Wei, ``On boundedness of error covariances
  for kalman consensus filtering problems,'' \emph{IEEE Trans. Autom. Control},
  vol.~65, no.~6, pp. 2654--2661, 2020.

\bibitem{6415998}
G.~N. Nair, ``A nonstochastic information theory for communication and state
  estimation,'' \emph{IEEE Trans. Autom. Control}, vol.~58, no.~6, pp.
  1497--1510, 2013.

\bibitem{cong2021rethinking}
Y.~Cong, X.~Wang, and X.~Zhou, ``Rethinking the mathematical framework and
  optimality of set-membership filtering,'' \emph{IEEE Trans. Autom. Control},
  vol.~67, no.~5, pp. 2544--2551, 2021.

\bibitem{moore2009introduction}
R.~E. Moore, R.~B. Kearfott, and M.~J. Cloud, \emph{Introduction to interval
  analysis}.\hskip 1em plus 0.5em minus 0.4em\relax SIAM, 2009.

\bibitem{8552364}
F.~Farina, A.~Garulli, and A.~Giannitrapani, ``Distributed interpolatory
  algorithms for set membership estimation,'' \emph{IEEE Trans. Autom.
  Control}, vol.~64, no.~9, pp. 3817--3822, 2019.

\bibitem{9591293}
------, ``An interpolatory algorithm for distributed set membership estimation
  in asynchronous networks,'' \emph{IEEE Trans. Autom. Control}, vol.~67,
  no.~10, pp. 5464--5470, 2022.

\bibitem{8114330}
X.~Ge, Q.-L. Han, and Z.~Wang, ``A dynamic event-triggered transmission scheme
  for distributed set-membership estimation over wireless sensor networks,''
  \emph{IEEE Trans. Cybern.}, vol.~49, no.~1, pp. 171--183, 2019.

\bibitem{Scott2016}
J.~K. Scott, D.~M. Raimondo, G.~R. Marseglia, and R.~D. Braatz, ``Constrained
  zonotopes: A new tool for set-based estimation and fault detection,''
  \emph{Automatica}, vol.~69, pp. 126--136, jul 2016.

\bibitem{9031321}
Y.~Cong, X.~Zhou, and R.~A. Kennedy, ``Finite blocklength entropy-achieving
  coding for linear system stabilization,'' \emph{IEEE Trans. Autom. Control},
  vol.~66, no.~1, pp. 153--167, 2021.

\bibitem{Chen1984}
C.-T. Chen, \emph{Linear system theory and design}.\hskip 1em plus 0.5em minus
  0.4em\relax Saunders college publishing, 1984.

\end{thebibliography}
\end{document}